%% file: bulges20pcnuclear.tex
\documentclass{emulateapj}
\usepackage{graphicx}
\def\r14{$R^{1/4}$}
\def\rn{$R^{1/n}$}
\def\kms{\ifmmode{\mathrm{km\,s}^{-1}}\else{km\,s$^{-1}$}\fi}
\def\kmsMpc{km\,s$^{-1}$\,Mpc$^{-1}$}
\def\HST{{\it HST}}
\def\reff{\ifmmode{R_{\rm e}}\else{$R_{\rm e}$}\fi}
\def\mueff{\ifmmode{\mu_{\rm e}}\else{$\mu_{\rm e}$}\fi}
\def\gammaSerPrime{\ifmmode{\gamma_{_{\rm S}}^{\,\prime}}\else{$\gamma_{_{\rm S}}^{\,\prime}$}\fi} 
\def\gammaNukPrime{\ifmmode{\gamma^{\,\prime}_{_{\rm N}}}\else{$\gamma^{\,\prime}_{_{\rm N}}$}\fi} 
\def\gammaNukAverage{\ifmmode{\langle \gamma\rangle}\else{$\langle \gamma\rangle$}\fi} 
\def\Pnull{$P_{\rm null}$}
\newcommand{\MassSun}{\ensuremath{{\cal M}_\odot}}
\newcommand{\MassBul}{\ensuremath{{\cal M}_\mathrm{Bul}}}
\newcommand{\MassBulDyn}{\ensuremath{{\cal M}_\mathrm{Bul,Dyn}}}
\newcommand{\MassGal}{\ensuremath{{\cal M}_\mathrm{Gal}}}
\newcommand{\MassCMO}{\ensuremath{{\cal M}_\mathrm{CMO}}}
\newcommand{\MassNuc}{\ensuremath{{\cal M}_\mathrm{Nuc}}}
\newcommand{\MassPt}{\ensuremath{{\cal M}_\mathrm{PS}}}
\newcommand{\MassSMBH}{\ensuremath{{\cal M}_\bullet}}

\newcommand{\MagKPt}{\ensuremath{M_{K,\mathrm{PS}}}}
\newcommand{\MagKExt}{\ensuremath{M_{K,\mathrm{Ext}}}}

\newcommand{\MagKBul}{\ensuremath{M_{K,\mathrm{Bul}}}}
\newcommand{\MagKSun}{\ensuremath{M_{K,\odot}}}
\newcommand{\LumKPt}{\ensuremath{L_{K,\mathrm{PS}}}}

\newcommand{\LumKBul}{\ensuremath{L_{K,\mathrm{Bul}}}}
\newcommand{\LumKSun}{\ensuremath{L_{K,\odot}}}

\shorttitle{Nuclei of bulges}
\shortauthors{Balcells, Graham, \& Peletier}

\begin{document}

\singlespace

\title{Galactic bulges from {\sl Hubble Space Telescope} NICMOS observations:  central galaxian objects, and nuclear profile slopes\altaffilmark{1},\altaffilmark{2}}
\author{Marc Balcells}
\affil{Instituto de Astrof\'\i sica de Canarias, 38200 La Laguna, Tenerife, Spain}
\author{Alister W. Graham\altaffilmark{3}}
\affil{Centre for Astrophysics and Supercomputing, Swinburne University of Technology, Hawthorn, Victoria 3122, Australia}
\and
\author{Reynier F. Peletier\altaffilmark{4}}
\affil{Kapteyn Institute, University of Groningen, 9700 AV Groningen, The Netherlands}
\email{balcells@iac.es}

\slugcomment{ApJ accepted 2007 april 27} 

\altaffiltext{1}{Based on observations made with the NASA/ESA {\sl 
Hubble Space Telescope}, obtained at the Space Telescope Science 
Institute, which is operated by the Association of Universities for 
Research in Astronomy, Inc., under NASA contract NAS 5-26555.}
\altaffiltext{2}{Based on observations made with the Isaac Newton Telescope operated on the island of La Palma by the Isaac Newton Group of Telescopes in the Spanish Observatorio del Roque de los Muchachos of the Instituto de Astrof\'\i sica de Canarias} 
\altaffiltext{3}{Also: Instituto de Astrof\'\i sica de Canarias, 38200 La Laguna, Tenerife, Spain}
\altaffiltext{4}{Also: School of Physics and Astronomy, University of Nottingham, NG7 2RD, UK}

\begin{abstract}
We have measured the central structural properties for a sample of S0-Sbc 
galaxies down to scales of $\sim$10~pc using {\it Hubble Space Telescope} 
NICMOS images.
We find that the photometric masses of the central star clusters, which 
occur in 58\% of our sample, are related to their host bulge masses such 
that $\MassPt = 10^{7.75\pm0.15}\,(\MassBul/10^{10}\MassSun)^{0.76\pm 
0.13}$.  Put together with recent data on bulges hosting supermassive 
black holes, we infer a {\it non-linear} dependency of the `Central 
Massive Object' mass on the host bulge mass such that $\MassCMO/\MassSun = 
10^{7.51\pm 0.06} (\MassBul/10^{10}\,\MassSun)^{0.84 \pm 0.06}$.  We argue 
that the linear relation presented by Ferrarese et al.\ is biased at 
the low-mass end by the inclusion of the disc light from lenticular 
galaxies in their sample.  
Matching our NICMOS data with wider-field, ground-based $K$-band images 
enabled us to sample from the nucleus to the disk-dominated region of each 
galaxy, and thus to perform a proper bulge-disk decomposition.  We found 
that the majority of our galaxies ($\sim$90\%) possess central light 
excesses which can be modeled with an inner exponential and/or an 
unresolved point source in the case of the nuclear star clusters.  All the 
extended nuclear components, with sizes of a few hundred pc, have disky 
isophotes, which suggest that they may be inner disks, rings, or bars; 
their colors are redder than those of the underlying bulge, arguing 
against a recent origin for their stellar populations.
Surface brightness profiles (of the total galaxy light, and the bulge 
component on its own) rise inward to the resolution limit of the data, 
with a continuous distribution of logarithmic slopes from the low values 
typical of dwarf ellipticals ($0.1 \leq \gamma \leq 0.3$) to the high 
values ($\gamma\sim 1$) typical of intermediate luminosity ellipticals; 
the nuclear slope bi-modality reported by others is not present in our 
sample.
\end{abstract}


\section{Introduction}
\label{Sec:Introduction}

The {\it Hubble Space Telescope (HST)} enables the study of 
the inner regions of nearby bulges and ellipticals down to
spatial scales of $\sim$10 pc, roughly one order of magnitude closer
to the center than is feasible with typical ground-based data.  These inner 
regions contain a small fraction of the ellipsoid mass, but they 
harbor the highest density regions of the galaxies and contain useful
clues to their formation.  

The availability of NIR array detectors in the nineties fostered significant progress in the understanding of many aspects of bulges, including structural parameters, colors, dust content and stellar populations, as well as the scaling of disk and bulge parameters, using ground-based imaging (e.g.\ Andredakis, Peletier, \& Balcells 1995, hereafter APB95; de Jong 1996; Seigar \& James 1998; Knapen et al.\ 1995; Khosroshahi et al.\ 2000;  Graham 2001a, hereafter G01; M\"ollenhoff \& Heidt 2001; Graham 2001b, 2002; Eskridge et al.\ 2002; MacArthur, Courteau, \& Holtzman 2003; Castro-Rodr\'\i guez \& Garz\'on 2003).    

NIR data helped to establish that exponential profiles provide better fits to the surface brightness profiles of bulges than \r14\ models (Kent et al.\ 1991 for the MW bulge; Andredakis \& Sanders 1994; de Jong 1996), and soon thereafter it was demonstrated that profiles of bulges of all Hubble types admit a particularly simple fit using the S\'ersic (1963; see Graham \& Driver 2005) function 

\begin{equation}
I(R) = I(0)\,\exp\{-b_n\,(R/\reff)^{1/n}\}
\label{Eqn:Sersic}
\end{equation}

\noindent(APB95; G01; M\"ollenhoff \& Heidt 2001; MacArthur et al.\ 2003; see Caon et al.~2003 for the case of elliptical galaxies). In eqn.~\ref{Eqn:Sersic}, \reff\ is the half-light radius of the bulge, and $b_n\approx 1.9992n-0.3271$.  The S\'ersic index $n$, which measures the curvature of the surface brightness profile, scales with bulge-to-disk luminosity ratio (B/D) and with bulge luminosity.   The S\'ersic index also provides a concentration parameter (Trujillo et al.\ 2001) which strongly correlates
with the velocity dispersion and central supermassive black hole mass (Graham et al.\ 2001a, 2001b), hence it is linked to global physical parameters of the
spheroid.  Numerical simulations also suggest that bulges have a range of profile shapes.  Aguerri, Balcells, \& Peletier (2001) show that the accretion of dense satellites onto disk-bulge-halo galaxies yields a growth of both the S\'ersic index and B/D, hinting that $n$ may be linked to the accretion history and to the growth of bulges.  $\Lambda$ cold dark matter cosmological simulations of galaxy formation yield bulge-disk structures where the bulge profile shape ranges from exponential to \r14 (Scannapieco \& Tissera 2003; Sommer-Larsen, G\"oth, \& Portinari 2003).  

The results given above, derived from ground-based data, bear the question of whether the inner regions to which the \HST\ gives access also follow the S\'ersic function.  Our picture of elliptical galaxy nuclei had to be revised in several ways after the \HST\  imaging campaigns.   Giant ellipticals often show a rather sudden inward flattening of their surface brightness profiles, confirming the result from ground-based data that some ellipticals have "cores" (Kormendy 1985), while intermediate-luminosity ellipticals ($-18\leq  M_B \leq -20.5$) do not show cores; their profiles approach power laws throughout the inner regions ("power-law" galaxies); see Faber et al.\ (1997, hereafter F97) and Rest et al.\ (2001, hereafter R01).  Inner profile slopes decrease toward fainter luminosities, and, for dwarf ellipticals, approach the slopes seen in the nuclei of giant, core galaxies, although dwarfs do not show profile discontinuities, i.e., do not show 'cores'  (Graham \& Guzm\'an 2003; Ferrarese et al.\ 2006, hereafter F06).  Many cores of ellipticals and S0s are dusty, and a fraction of them harbor central unresolved sources at \HST\  resolution (Lauer et al.\ 1995, hereafter L95; Phillips et al.\ 1996; Carollo et al.~1997; Ravindranath et al.\ 2001; Stiavelli et al.\ 2001).  
Inasmuch as bulges share global similarities with ellipticals when studied from the ground, we enquire whether bulges show "cores", whether bulges show nuclear sources.  

Bulges of disk galaxies have been targeted less often than ellipticals by the \HST. 
Peletier et al.\ (1999, hereafter Paper\,I) analyzed a sample of 19 field S0-Sbc galaxies using WFPC2 F450W, F814W and NICMOS F160W images, with the goal of obtaining bulge stellar population diagnostics.  The combination of blue and NIR colors allowed them to put tight limits on the ages of bulge populations.  Ages of S0 to Sb bulges were found to be comparable to those of ellipticals in the Coma cluster, with a small age spread $<2$ Gyr (Sbc bulges showed colors corresponding to younger ages).   Nuclei were found to be dusty, with $A_V =0.6-1.0$ mag.  

Carollo and collaborators surveyed mid- to late-type bulges using WFPC2 and NICMOS (e.g.\ Carollo 1999; Carollo \& Stiavelli 1998; Carollo et al.\ 1997, 1998, 2001, 2002; Seigar et al.\ 2002).  These authors focus on bulge structure.  They provide fits using the \r14, exponential and Nuker models, and propose a structural classification of bulges into `\r14
classical' and 'exponential'.
Carollo et al.\ (2002) find nuclear resolved components (NC) in the centers of 60\% of the exponential bulges.  In their view, '\r14' and 'exponential' bulges respectively show 'high' and 'low' nuclear profile slopes, a structural difference which would trace different formation histories.  

Whether bulges come in two families with distinct structural properties has implications for formation mechanisms of bulges.  Several models have been proposed (see Wyse, Gilmore, \& Franx 1997; Bouwens, Cayon, \& Silk 1999; Kormendy \& Kennicutt 2004): early collapse (Renzini 1999; Zoccali et al.~2003); mergers prior to disk formation (Kauffmann, Charlot \& White 1996); satellite accretion (Pfenniger 1993; Aguerri et al.\ 2001); and disk instabilities (Pfenniger \& Norman 1990; Zhang 1999).  
Bulges with \r14\ structure fit in the early collapse or merger scenarios, while exponential bulges are destroyed by mergers (Aguerri et al.\  2001) and may instead be expected from disk instabilities (Combes et al.\ 1990).  
Edge-on, peanut-shaped bulges are known to have bar dynamics, and are therefore also expected to form from disk instabilities (Kuijken \& Merrifield 1995; Bureau \& Freeman 1999).  
The existence of two classes of bulges is commonly understood as evidence that massive bulges come from mergers while less massive bulges grow as a result of disk instabilities (see, e.g., Athanassoula 2005).  

In this paper we analyze the structural properties of bulges of early- to intermediate-type galaxies at \HST\ resolution using the S0-Sbc sample presented in Paper I.  We address profile shapes, nuclear sources, nuclear slopes, and central massive black hole mass estimates.  Given the ability of the S\'ersic model to describe the profiles of spheroids at ground-based resolution, we use the S\'ersic model as our starting point and enquire whether the increased spatial resolution of the \HST\  contributes to support or to modify the ground-based picture.   We perform a bulge-disk decomposition of the surface brightness profiles using combined \HST+ground-based profiles that sample the galaxy light distribution from the nucleus to the disk-dominated region.  Ignoring this step would bring up two problems: the un-modeled disk contribution to the inner profile would bias the bulge nuclear parameters; and, we would not be able to derive basic bulge parameters such as the total luminosity and the effective radius as the \HST\  images do not cover the entire bulge at the distances of our target galaxies. 

We avoid using the \r14\ or exponential models, rather we focus on S\'ersic fits to the bulge profiles to test if the profile shape dichotomy appears when it is not forced.  Our first results on bulge profile shapes using \HST\  data were presented in Balcells et al.\ (2003, hereafter Paper\,II).  In that paper we show that \r14\ bulge profiles are exceedingly rare.  In this and a companion paper (Balcells, Graham, \& Peletier 2007, hereafter Paper~IV) we perform a comprehensive analysis of those profiles.  We will show that inner surface brightness profiles show excesses, over the best-fit bulge S\'ersic model, which can be successfully modeled by adding central unresolved sources and/or inner exponential components to the fitting function (\S\,\ref{Sec:BulgeDiskFits}).  Section~\ref{Sec:ParameterUncertainties} provides details on the estimation of  parameter errors through fits to simulated profiles.  The subsequent sections analyze the properties of the nuclear excess light.  Sect.\,\ref{Sec:NuclearExtendedComponents} shows that the galaxies with extended nuclear components closely match those with nuclear disky isophotes, which suggests that the excess light in the surface brightness profiles comes from flattened components such as disks, rings or inner bars.  Sect.\,\ref{Sec:PointSources} derives luminosities and masses for the unresolved nuclear sources and addresses the Compact Massive Object (CMOs) paradigm, i.e., that nuclear star clusters are the low-mass extension to central supermassive black holes.  In \S\,\ref{Sec:SMBH} we relate the point sources to black hole mass estimates from the bulge velocity dispersions.  Finally, in \S\,\ref{Sec:Gammas} we present and discuss the nuclear surface brightness profile slopes and compare them to those of ellipticals, bulges and dwarf ellipticals.  
In Paper~IV, we discuss global bulge and disk scaling relations as inferred from the profile decompositions.  
A Hubble constant of $H_{0} = 75$ km\,s$^{-1}$\,Mpc$^{-1}$ is used 
throughout.  

\section{Galaxy sample and data}
\label{Sec:Data}

We have analyzed 19 galaxies from the Balcells \& Peletier (1994, hereafter BP94)
diameter-limited sample of inclined, early-to-intermediate type disk
galaxies.  The BP94 sample was selected from the Uppsala General Catalog of Galaxies (UGC; Nilson 1973) to include all disk galaxies of types S0 to Sbc,  listed as unbarred in the UGC, with blue diameters greater than 2 arcmin, inclinations above 50$^\circ$ (i.e.\ mid- to high-inclination), and apparent blue magnitudes brighter than 14.0 mag, within given limits of equatorial and Galactic coordinates.  Upon inspection, some cases were excluded due to being obviously barred, interacting, or very dusty, leaving 30 galaxies that were analyzed in BP94, APB95, and Peletier \& Balcells (1996, 1997).  The present subsample comprises 19 galaxies of types S0 to Sbc that were imaged with \HST\  (NICMOS-F160W [camera 2] and WFPC2-F450W and F814W; Paper\,I).  The subsample was selected to provide representative examples of each Hubble type, and to exclude cases where dust obscured the nuclei.  Due to their high inclinations, some of the galaxies may harbor bars which go undetected in the images; bars may be suspected from the peanut-shaped isophotes of some of the bulges (e.g.\ Kuijken \& Merrifield 1995).  None of the galaxies have a Seyfert or a starburst nucleus.   

The sample has been extensively studied by us in previous papers.  
Nuclear colors at \HST\  resolution and bulge ages have been presented in Paper\,I, where postage-stamp images of the \HST\  data for the 19 galaxies studied here may be found.  
Peletier \& Balcells (1997) published $K$-band surface brightness profiles and isophotal parameters from ellipse fits to wider field-of-view UKIRT images.  Central stellar  velocity dispersions and a Fundamental Plane analysis are given in Falc\'on-Barroso, Peletier \& Balcells (2002).  Minor-axis kinematic profiles are presented in Falc\'on-Barroso et al.\ (2003).   

The inclined viewing angle for this sample presents advantages and disadvantages for a structural study of the nuclear properties of disk galaxies.  The main drawback is extinction, which can completely hamper detection of nuclear structures at visible wavelengths.  Fortunately, extinction is smaller at NIR wavelengths.  Paper~I concludes that extinction for this sample is on average $A_{H}=0.1 - 0.2$ mag in the nucleus, and much lower further out.  Problems related to extinction should therefore be minor for the study presented here.  The main advantage of working with an inclined sample is that isophotes provide information on the flatness of each galaxian component, and hence, they guide in the identification of these components.  

Basic properties of the sample are given in Table~\ref{Tab:Sample}, where we list distances, spatial scales, $K$- and $R$-band absolute magnitudes, central velocity dispersions, and disk ellipticities.  Distances are derived from recession velocities relative to the Galactic-standard of rest, from de Vaucouleurs et al. (1991, hereafter RC3).  
We present new $K$-band apparent and absolute magnitudes for the program galaxies. The photometry we had published in APB95 is about 0.5 mag brighter than that from 2MASS; new ellipse fits to the APB95 images yield total $K$-band apparent magnitudes that closely match 2MASS total magnitudes (mean difference of 0.00 mag, rms of 0.21 mag).   The coincidence of our photometry and that of 2MASS makes us  believe that the present photometric zero points are more accurate than those of APB95.   Therefore we adopt the new apparent magnitudes in this paper, except for NGC~5746 and 5965, which overfill our frames, and for NGC~5879, whose frame suffers from bad sky planarity.  For those three galaxies we adopt the total absolute magnitudes from NED.  Adopted apparent magnitudes are listed in Table~\ref{Tab:Sample}.  Absolute $K$-band magnitudes are corrected for Galactic extinction, cosmological dimming, and K-correction.  Errors in the $K$-band absolute magnitude include the photometric error and a distance modulus error which assumes a flat 50 \kms\ recession velocity error.  
We use the disk ellipticities from APB95, which have been derived on $K$-band images;  these ellipticities show minimal differences, of at most 0.05, with respect to the $R$-band values given in BP94.  

\input tabsample.tex
\input tabfitpars.tex
\input tabmainphyspars.tex
\input tabnucphyspars.tex

Details of the \HST\  observations and the reduction of the \HST\  data are given in Paper I.  
Here we derive elliptically-averaged surface brightness profiles and
isophotal shapes from the HST/NICMOS F160W images (19\arcsec$\times$19\arcsec, 0.075 arcsec/pixel), 
from 0.03 arcsec (1/2 a pixel size) to typically 10 arcsec, using the
{\tt galphot} package (J{\o}rgensen, Franx, \& Kj{\ae}rgaard 1992). We keep the centers fixed at the galaxy luminosity peak, and let the ellipticity and position angle of the isophotes vary.    

To extend the surface brightness profiles to large radii, we use the elliptically-averaged $K$-band surface brightness
profiles derived from UKIRT/IRCAM3 images (mosaics of
75\arcsec$\times$75\arcsec\ frames, 0.291\arcsec/pixel) published by Peletier \&
Balcells (1997), which we transform to $H$-band by approximating the
$H-K$ profiles with the transformation

\begin{equation}
H-K = 0.111(I-K) - 0.0339
\end{equation}

\noindent derived from the GISSEL96 models of Bruzual \& Charlot (see Leitherer et al.\  1996) using $I-K$ profiles from Peletier \&
Balcells (1997).  

The \HST\  and ground-based (GB) profiles have matching
slopes in the range $3''\leq r \leq 8''$ and show  zero-point
offsets that are always below 0.1 mag.  We correct these by applying
an offset to the GB profiles, which overall have less photometric
accuracy.  The process described here is the same that was employed in Paper~I
to derive color profiles for bulges at \HST\  resolution.  The match of GB to \HST\ profiles is extremely good, as can be seen in the residual profiles presented below (see Appendix, Fig.\,\ref{Fig:MuSigColGeom}).  The resulting geometric-mean-axis profiles were used in Paper~II for the analysis of the S\'ersic shape index $n$.  

\section{Profile decomposition}
\label{Sec:BulgeDiskFits}

The decomposition of the surface brightness profiles was described in Paper\,II.  
Briefly, the combined \HST+GB profiles were fitted with a
PSF-convolved S\'ersic plus exponential law using the code described
in G01, modified to use a Moffat PSF ($\beta=6.9$).  The Moffat FWHM=0.131\arcsec\ used here is slightly narrower than that used in Paper\,II. This has two consequences: first the light in the NICMOS PSF wings needs to be accounted for with a suitable correction 
(\S~\ref{Sec:ParameterUncertainties}); second, more nuclear components appear as resolved in the present fits than they did in Paper~II.  
Fitting proceeded by chi-square minimization, allowing all 5 free parameters (disk $\mu_{0}, h$; bulge \mueff, \reff, $n$) to vary.  Convergence was generally straightforward, except for NGC~5326 and NGC~ 5854, where we fixed the disk parameters by eye.  In a few other cases, a few outer los S/N points were excluded to prevent downward deviations from distorting the fit in the main part of the disk (see Figure\,\ref{Fig:Profiles}).

We found that pure S\'ersic+exponential fits to the entire radial range of the profiles provide an inaccurate description of the data: the residuals from the fits show a strong wave pattern with an obvious central positive residual.  An example of this feature is shown in Figure~1 of Paper~II.  
As a consequence, the bulge and disk fitted parameters 
are highly sensitive to any inner radius cutoff imposed on the fitting range.  
Such fits yield uncertain values for the bulge S\'ersic index $n$, the total bulge luminosities, and the bulge-to-disk ratios.
Fits excluding the inner $\sim$0.5--1.0 arcsec in radius (a common approach for deriving bulge parameters in the presence of nuclear components, e.g., Carollo et al.\ 1998; Stiavelli et al.\ 2001) 
show strong central residuals (see Fig.\,\ref{Fig:MuSigColGeom}, second row).  These indicate that our galaxy surface brightness profiles cannot be modeled with just the sum of a S\'ersic bulge and an exponential disk. 
For 14 out of 19 galaxies, we find an excess of central light above the S\'ersic bulge; 2 galaxies show central
depressions, while 3 follow the S\'ersic profile reasonably well over the entire radial range.  These numbers do not substantially vary when modifying the inner radial cutoff. 

The failure of pure S\'ersic+exponential fits suggests that many of our program galaxies harbor nuclear components.  The detection of central excesses has previously been reported for bulges, dwarf ellipticals, and intermediate-luminosity ellipticals (Aaronson 1978; Binggeli, Sandage \& Tarenghi 1984; Caldwell \& Bothun 1987; Phillips et al.\ 1996; Carollo et al.\ 1997; R01; Ravindranath et al.\ 2001).  Excluding the central $\sim$1\arcsec\ has been a common strategy to cope with these central components, which also avoids the problems with central dust extinction, e.g.\ the Stiavelli et al.\ (2001) fits to \HST-based dwarf elliptical surface brightness profiles, or the \r14\ or exponential fits to bulge surface brightness profiles by Carollo, Stiavelli, \& Mack (1998).  In our case, outside 1\arcsec, the pure S\'ersic+exponential bulge-disk fits are quite satisfactory (see Figure\,\ref{Fig:MuSigColGeom}), suggesting that the S\'ersic model provides a good approximation to the large-scale brightness profiles of bulges.  


We therefore assume that the S\'ersic model describes "the bulge", taken as the spheroidal component residing in the center of the disk galaxy, and that any central excess above the S\'ersic profile is due to additional photometric components.  
To implement our approach, we run the fitting program with additional central components.  We tested central point sources, central exponential disks, a combination of both, and central Gaussians of free width.  In all cases, the models were convolved with the Moffat PSF prior to fitting.   For each galaxy, all fits were inspected, and we selected the fits which, with a minimum number of added components, resulted in residual profiles without structures such as the wave pattern described above.  In all cases, 
we adopted the extra component when the root-mean-square (rms) of the residuals improved by over 10\%, and we discarded it when the rms improved by less than 10\%.  
From a formal point of view, this approach is justified given that, with typically 90 independent data points, adding a point source to the 5-parameter S\'ersic+exponential fit would trivially yield an rms decrease of $<2.8\%$; including an additional nuclear exponential to the above solution would trivially improve the rms of that solution by 4.0\%; and,  
an rms trivial improvement of 10\% would be expected only after adding 18 constraints to the fit.  Moreover, the adequacy of the employed model is ascertained through examination of the shape of the residual profile.  
Fits with residual profiles that were featureless were adopted without testing more complex models, as the rms cannot be improved by modifying the functional form of the fitted model in those cases.  
Residual profiles for the adopted solutions have $0.02 \leq {\rm rms} \leq 0.1$ magnitudes.  

Hence, our approach has two steps: first, a $\chi^2$ minimization of five different models, namely: pure S\'ersic+exponential; S\'ersic+exponential+Moffat; S\'er\-sic+expo\-nen\-tial+inner expo\-nen\-tial; S\'er\-sic+expo\-nen\-tial+Moffat+in\-ner expo\-nen\-tial; and S\'er\-sic+expo\-nen\-tial+Gaussian.  Second, a selection of the best of these models, applying Occam's razor to choose the simplest model which describes the data.  Our solutions describe the surface brightness profiles to a high degree of accuracy, although we generally cannot guarantee  to have obtained unique solutions.  Parameter uncertainties are discussed in \S~\ref{Sec:ParameterUncertainties}.  


Figure\,\ref{Fig:Profiles} shows the profiles, the best-fit models, and the residual profiles from the fits for the 19 galaxies.  Parameters for the fits are given in Table~\ref{Tab:FitParameters}.  
The largest structures in the residual profiles occur in the region of the disk, and are due to spiral arms and other disk features.  In the bulge-dominated region, the small oscillation around 0.2 arcsec arises from the cross-pattern in the NICMOS PSF wings and deviations from the assumed Moffat model (\S~\ref{Sec:ParameterUncertainties}).  

Internal extinction is obviously an issue at the galaxy centers.  Fortunately, our use of NIR imaging data strongly alleviates the problem.  Paper~I shows that $A_{V,{\rm Center}} = 0.6 - 1.0$ mag on average for this sample, which corresponds to $A_{H}$ between 0.1 and 0.2 mag.  Extinction is highly concentrated in the inner $\sim$100 pc, hence is probably local rather than due to intervening disk dust.  This extinction must vary from galaxy to galaxy, and can be estimated from the excess of central $I-H$ color with respect to a typical bulge stellar population.  The resulting values are so small that we do not apply extinction corrections to our photometry.  The lack of such correction does not affect any of the conclusions of the paper.  

\begin{figure*}[htbp]
\begin{center}
\includegraphics[height=15cm]{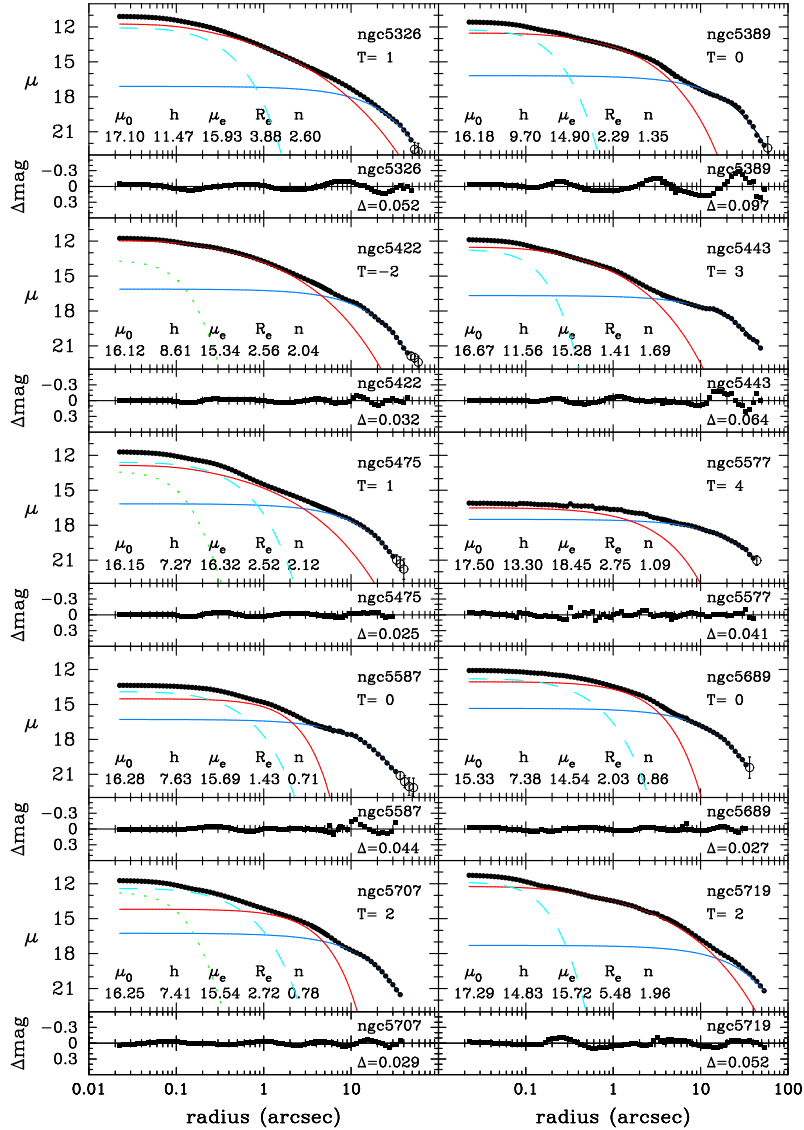}
\caption{$H$-band, combined \HST\  plus ground-based surface brightness profiles for the 19 galaxies, and best-fit models. The abscissa is the geometric mean radius in arcsec, $r \equiv a\,\sqrt(1-\epsilon)$, $a\equiv $ major axis, $\epsilon \equiv $ ellipticity.  {\it Filled circles:} fitted data points.  {\it Open circles:}  outer points excluded from the fit.  {\it Solid lines:}  seeing-convolved S\'ersic bulge and exponential outer disk models.  {\it Dotted lines:} seeing-convolved inner point sources.  {\it Dashed lines:} inner exponential components.  Given in each panel are the best-fit outer disk parameters (central surface brightness $\mu_0$, disk scale length $h$) and bulge parameters (effective surface brightness $\mu_e$, effective radius $R_e$ and S\'ersic index $n$).  Parameters for the inner components are given in Table~\ref{Tab:FitParameters}. Below each profile is the profile residual (data {\it minus} model); $\Delta$ gives the root-mean-square residual from the fit, in magnitudes.  \label{Fig:Profiles}}
\end{center}
\end{figure*}

\addtocounter{figure}{-1}

\begin{figure*}[htbp]
\begin{center}
\includegraphics[height=15cm]{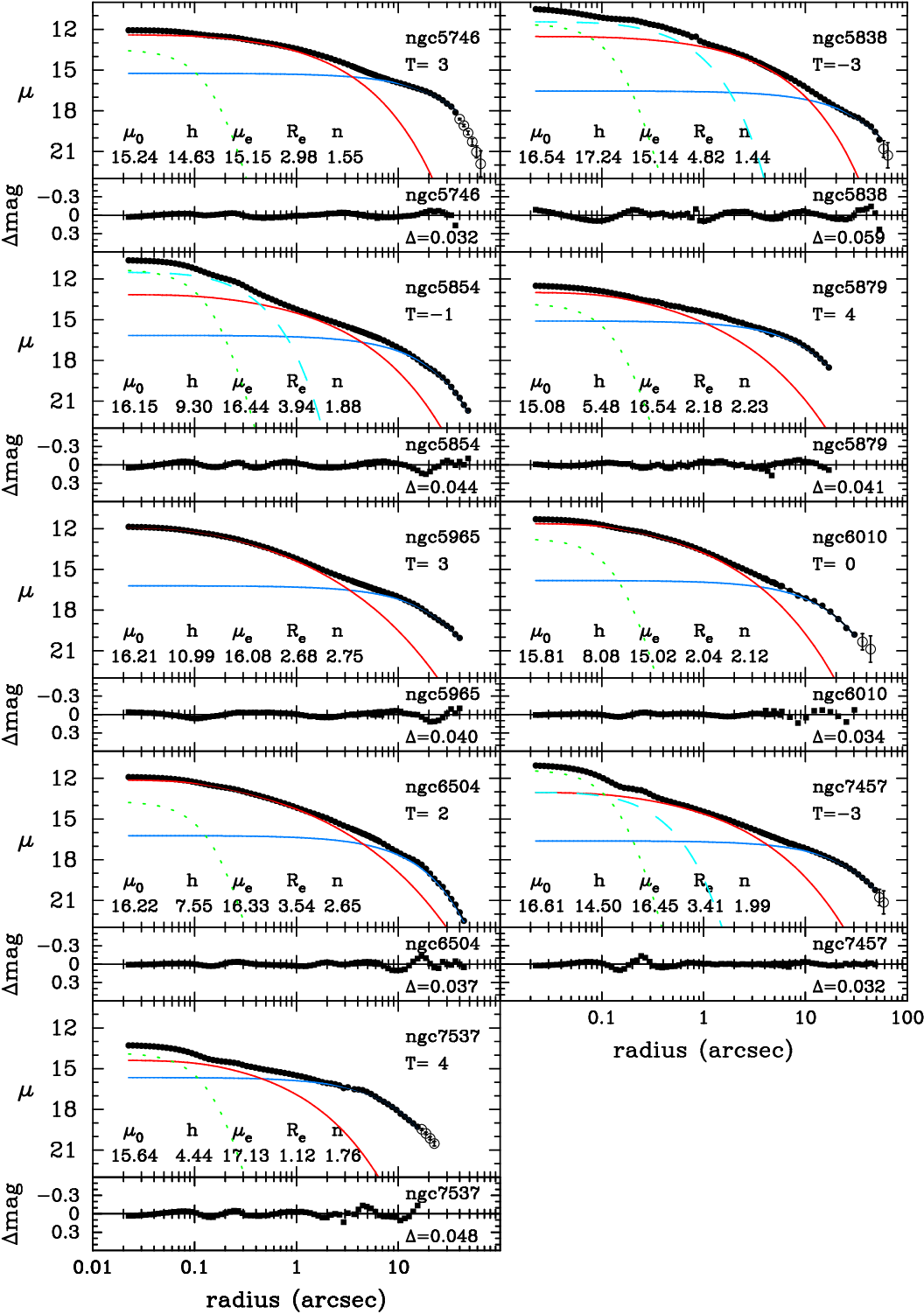}
\caption{{\it continued.}}
\label{default}
\end{center}
\end{figure*}

\begin{table}[htdp]
\caption{Parameter uncertainties}
\begin{center}
\begin{tabular}{lrrrrr}
\hline\hline
	& \multicolumn{2}{c}{\em se,pse models} & 
	& \multicolumn{2}{c}{\em ese models} \\ \cline{2-3} \cline{5-6}
Param		& \multicolumn{1}{c}{$\Delta$}	
			& \multicolumn{1}{c}{$\sigma$} & 
			& \multicolumn{1}{c}{$\Delta$}
			& \multicolumn{1}{c}{$\sigma$} \\
(1) 	& \multicolumn{1}{c}{(2)} & \multicolumn{1}{c}{(3)} 
&	& \multicolumn{1}{c}{(4)} & \multicolumn{1}{c}{(5)} \\ \hline
Outer exponential (disk): & & & & \\ 
$\mu_0$		& 0.089		& 0.123 &	& -0.010	& 0.036		\\
$\log(h)$		& 0.007		& 0.014 && -0.003	& 0.004		\\ 
Mag			& 0.029		     & 0.085 &&  0.010	& 0.046              \\ 
$-2.5\,\log(D/T)$	& 0.066	& 0.101	&& 0.012 & 0.046  \\ \hline
S\'ersic (bulge): & & & & \\ 
\mueff		& 0.060		& 0.224 && 0.205	& 0.348		\\ 
$\log(\reff)$ 	& 0.029		& 0.052 && 0.044	& 0.070		\\
$\log(n)$		& -0.007		& 0.035 && 0.007	& 0.072		\\ 
Mag			& -0.090		     & 0.106 &&  -0.019	& 0.150              \\
$-2.5\,\log(B/D)$	& -0.118	& 0.159	&& -0.025 & 0.163  \\
$-2.5\,\log(B/T)$	& -0.052	& 0.071	&& -0.017 & 0.118  \\ \hline
Nuclear unresolved source: & & & & \\ 
Mag$_\mathrm{PS}$	& 0.310		& 0.351  &&		&		\\ \hline 
Nuclear exponential, {\it brigth range}: & & & & \\ 
$\mu_{0,2}$	&  			& 	 	& 	& 0.45	& 0.45		\\
$\log(h_2)$	& 			&		&	& 0.09	& 0.04		\\
Mag$_2$		& 			& 		&	& -0.003	& 0.26 \\ 
Nuclear exponential, {\it faint range}: & & & & \\ 
$\mu_{0,2}$	   &  		    & 	 	& 	& 0.81	& 0.94           \\
$\log(h_2)$	& 			&			&	& 0.18	& 0.18\\
Mag$_2$		& 			& 		&	& -0.09	& 0.47\\ \hline
\end{tabular}
\end{center}
\label{Tab:Simulations}
\tablecomments{For each parameter, cols. (2) and (3) give the mean difference (measured minus input) and rms deviation for \textit{se, pse} synthetic models.  Columns (3) and (4) give the same quantities for \textit{ese} models.  For the nuclear exponentials, \textit{bright range} (\textit{faint range}) denotes models in which the inner exponential is brighter(fainter) than the S\'ersic component at $R=0.1\arcsec$.  See \S~\ref{Sec:ParameterUncertainties}.}
\end{table}%

\section{Parameter uncertainties}
\label{Sec:ParameterUncertainties}

Simulations with synthetic profiles were carried out to estimate errors in parameter recovery.  
The simulation work addresses two main questions.  First is parameter coupling, especially between bulge and nuclear components, e.g., a profile consisting of a high-$n$ S\'ersic and an outer exponential might be reproduced by the fitting program as a nuclear component, a lower-$n$ S\'ersic and an outer exponential.  
A second important issue is whether our choice of a Moffat PSF is adequate.  The Tiny-Tim PSF for our NICMOS images comprises a central peak surrounded by a secondary maximum at about 0.23\arcsec.  An analytical PSF such as the Moffat function is particularly convenient for one-dimensional profile fitting, but may affect the parameters derived for nuclear sources, and the light in the secondary maximum may masquerade as an extended component.  

For the simulations, synthetic images were generated with IRAF's {\tt mkobjects} task, comprising a S\'ersic component, an outer exponential component, with and without nuclear components.  Nuclear components were either a point source; a Gaussian; or an inner exponential.  The sampled range for the parameters was bigger than that displayed by our target galaxies.  The images were convolved with Tiny-Tim PSFs derived from the HST NICMOS images of the target galaxies, and noise was added to yield surface brightess profile errors similar to those of the target galaxies.  Surface brightness profiles were derived for each simulated galaxy image by fitting ellipses, using the same fitting parameters that were employed for the derivation of the profiles for the program galaxies. The radial extent of the profiles was somewhat lower than that of the program galaxies, but this should not affect the results given that, as we show below, disk parameters were accurately reproduced with their current radial extent.  
The profiles were fitted with combinations of S\'ersic, exponential and nuclear components using the same code used for the program galaxies.  As a test for any tendency of the fitting code to add non-existent nuclear components, all models were fitted with and without nuclear components in the fitting function.  

We found that nuclear Gaussian components are particularly difficult to reproduce.  We suspect that the quadratic dependence of the Gaussian function on $r$ made the fits unstable.  Because of the failure with synthetic profiles, we do not present fits to the program galaxies employing Gaussian nuclear components.  We note however that such fits yielded results that were generally consistent with the results of PS or exponential nuclear components: galaxies well modeled with inner PSs yielded good fits with very narrow Gaussians, and galaxies that required an inner exponential component yielded good fits with broad Gaussians as well.  

Results from the simulations are summarized in Table~\ref{Tab:Simulations}, which lists mean offsets (measured \textit{minus} input) and rms deviations for each group of models.  
For the statistics, we group together pure S\'ersic+exponential ({\it se}) models and PS+S\'ersic+exponential ({\it pse}) models, which show similar uncertainties, and list models with inner exponentials ({\it ese}) separately; differences between the two sets are small anyway.  Disk parameters are recovered with high accuracy.  Uncertainties for bulge parameters are only somewhat larger, and are asymmetric: we tend to recover fainter \mueff\ (0.2-0.3 mag\,arcsec$^{-2}$), and larger \reff\  ($\Delta(\log(\reff))\sim 0.05$).  Such average uncertainties are probably overestimates, as the simulated models include inner components more luminous and more extended than seen in our program galaxies.  

Point sources are recovered with an offset of $\sim$0.3 mag.  Most likely, this offset arises from the fractional light in the secondary maximum of the NICMOS PSF (0.5 mag outside 2 pixels for F160W; Holfeltz \& Calzetti 1999).  
The width of our Moffat PSF (FWHM = 0.131\arcsec) was set to match the main peak of the NICMOS PSF, hence the light in the wings gets unaccounted for; a MOFFAT width of $\sim$0.19\arcsec\ would greatly reduce such offset, but at the price of losing spatial resolution, therefore we adopt the narrower PSF and simply apply a 0.31 mag aperture correction to the point source magnitudes.  

For inner exponentials, models with $\mu_{0,2}$  $>$1 mag fainter than the S\'ersic's $\mu_0({\rm Sersic})$ get lost in the noise, while, at the bright end, models with $\mu_{0,2} - \mu_0({\rm Sersic}) < -2$ yield inner profiles dominated by the inner exponential, which are quite unlike those of real galaxies.  The statistics shown in Table~\ref{Tab:Simulations} correspond to the range between those limits, which we split in bright and faint ranges as shown in the Notes to the table.  The PSF wings affect inner exponential components by making their derived scale-lengths about 10-20\% higher, while making the central surface brightness fainter.  Total magnitudes show an uncertainty of a few tenths of magnitude -- larger for the fainter components as expected.  

Three main lessons derive from the simulations.  ($i$) Bulge and outer disk parameters are robust to the presence of nuclear components and to the choice of PSF; in particular, our fitting code does not artificially introduce nuclear components to pure S\'ersic+exponential profiles with a high-$n$ S\'ersic component.  ($ii$) A moderate level of parameter coupling is present for the bulge: derived \mueff\ are probably faint by $\sim$0.2-0.3 mag\,arcsec$^{-2}$, while $\log(\reff)$ are probably overestimated by $\sim$5\%. ($iii$) Nuclear unresolved sources are accurately recovered once an aperture correction of $\sim$0.3 mag is applied. 

\section{Results}
\label{Sec:Results}

Physical parameters for bulges, disks, and nuclear components are listed in Tables~\ref{Tab:MainPhysParameters} and \ref{Tab:NucPhysParameters}.  Corrections for measurement offsets, derived from the simulations (\S\ref{Sec:ParameterUncertainties}), have been added where applicable (Table~\ref{Tab:Simulations}).  RMS error estimates derive as well from Table~\ref{Tab:Simulations}.  

The first result of our analysis is that nuclear sources are quite common in disk galaxy bulges.  
Excesses over the S\'ersic fit to the bulge are measured in 17 out of 19 galaxies.   Over half (11) of those are extended; six can be modeled with an exponential profile, while five harbor a central unresolved source in addition to the exponential.  
The remaining 6 galaxies with central excesses can be modeled with an unresolved source.  
These results are not new.  Nuclear sources, resolved or unresolved at the scale of the \HST\ instruments have been reported by others, for bulges (Carollo et al.\ 1998; Paper I), for ellipticals and early-type bulges (R01; Ravindranath et al.\ 2001; F06), for dwarf ellipticals (Stiavelli et al.\ 2001; Graham \& Guzm\'an 2003; C\^ot\'e et al.\ 2006), and for late-type spirals (B\"oker et al.\ 2002).  The variety of detection methods and functions used by these teams to model the underlying light distributions suggests that nuclear components are not an artifact of the model fitting but are real components of many galaxy nuclei.  

\begin{figure*}[htbp]
\begin{center}
\includegraphics[height=7cm]{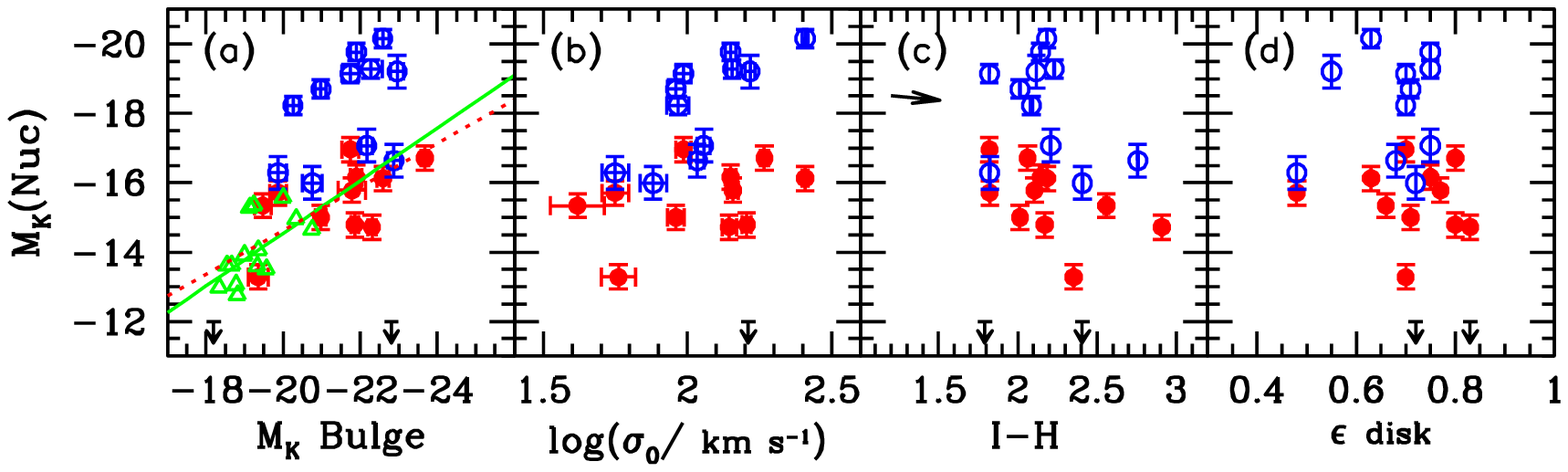}
\caption{\label{Fig:PSvsMK} Absolute $K$-band magnitude of the nuclear components against: ({\it a}) the $K$-band absolute magnitude of the bulge.  {\it Filled circles}: unresolved components in bulges.  {\it Open circles}: resolved components in bulges.  {\it Triangles}: nuclear components in dwarf ellipticals from Graham \& Guzm\'an (2003).  
The \textit{dotted line} is an orthogonal regression to the bulge unresolved components, see eqn.~\ref{Eqn:PSvsMK}, while the \textit{solid line} is an orthogonal regression to the bulge and dE unresolved components together (eqn.~\ref{Eqn:PSdEvsMK}).  
({\it b}) aperture-corrected central velocity dispersion, from Table\,\ref{Tab:Sample}.  
({\it c}) Central $I-H$ color, from HST/NICMOS F160W and WFPC2 F814W images (Paper\,I).  An extinction vector for normal Galactic extinction of $A_{V}=1$ mag (Rieke \& Lebofsky 1985) is plotted in the top-left.  ({\it d}) Ellipticity of the outer disk, from two-dimensional bulge-disk decomposition in $K$-band images (APB95).  Upper limits are given for the galaxies without detected nuclear components; one of those (NGC~5577) does not have a central velocity dispersion measurement.}
\end{center}
\end{figure*}

Absolute magnitudes of the nuclei are plotted in Figure~\ref{Fig:PSvsMK} against $K$-band absolute magnitude of the bulge, bulge central velocity dispersion, bulge color and disk ellipticity.  
Nuclei are on the mean $4.4 \pm 1.9$ mag fainter than their host bulges, and $6.1 \pm 1.7$ mag fainter than their host galaxies, although some of the extended components are only 2 mag fainter than their host bulge (e.g., NGC~5587, see Table~\ref{Tab:NucPhysParameters}).  In general, nuclei are minor contributors to the galaxy light.





If structural components are linked to formation events in the history of the host galaxies, we may enquire whether  nuclei are recent additions to the galaxy, or are they old, perhaps the seed of the formation of the bulge or of the entire galaxy.  If galaxy centers host a supermassive black hole, whose formation and growth went through a phase of positive feedback with the formation of the galaxy (Silk 2005), are the nuclei detected here connected in any way to such positive feedback (McLaughlin et al.\ 2006)?  
In the following subsections we further analyze the extended and unresolved nuclear  components.  

\begin{figure}[htbp]
\begin{center}
\includegraphics[width=0.45\textwidth]{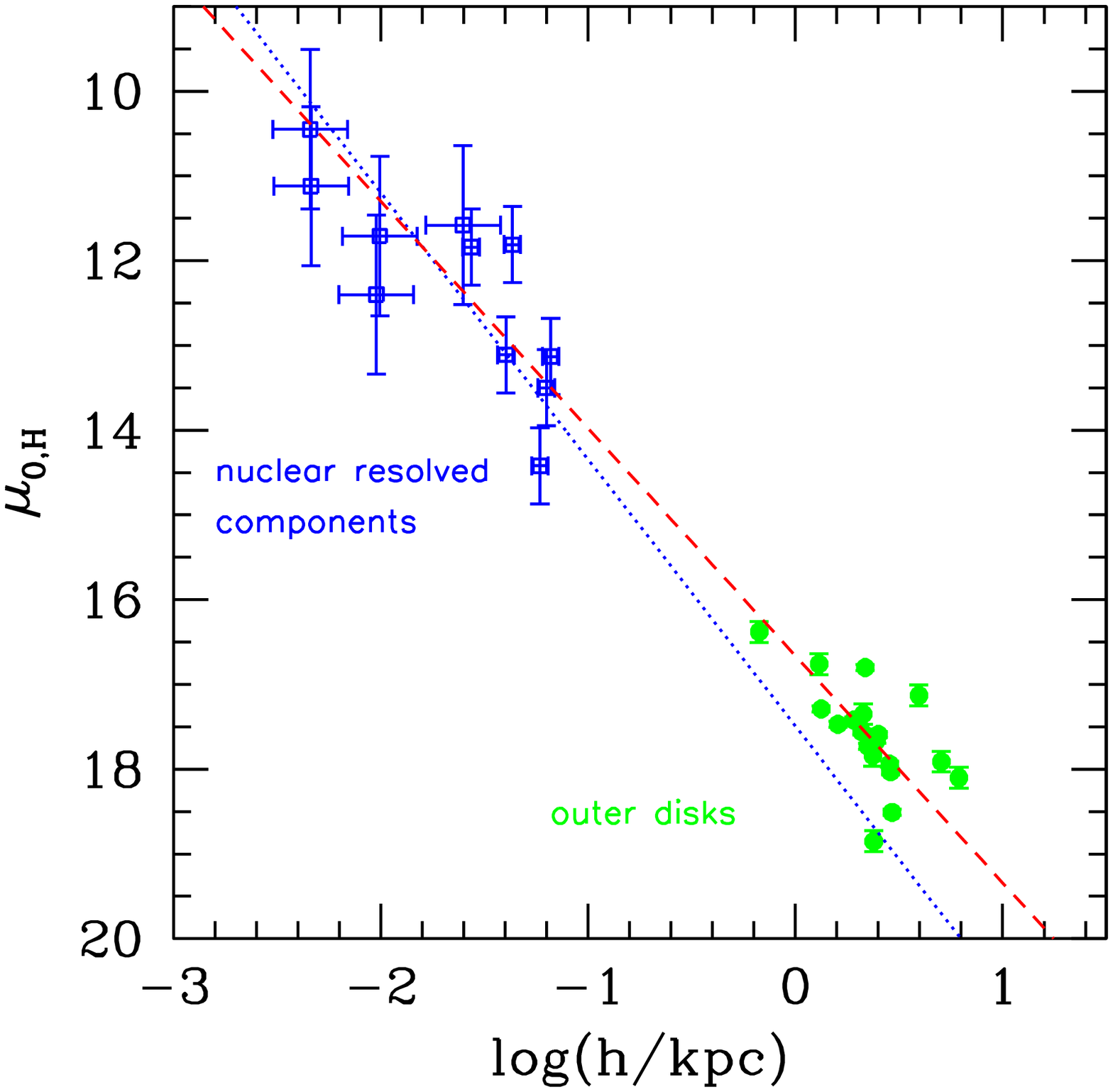}
\caption{Disk scale-length in kpc vs.\ extrapolated central surface brightness, for inner extended components, and for main galaxy disks.  \textit{Dotted line:} orthogonal regression to the nuclear resolved components (Eqn.~\ref{Eqn:Mu0vsHinner}).  
\textit{Dashed line:} orthogonal regression to both nuclear resolved components and outer disks (Eqn.~\ref{Eqn:Mu0vsHall}).  }
\label{Fig:mu0vsh}
\end{center}
\end{figure}

\subsection{Nuclear extended components}
\label{Sec:NuclearExtendedComponents}

The nuclear extended components have absolute $K$-band magnitudes in the range $-16 > \MagKExt < -20$.  Their scale lengths are $\sim$5 to 60 pc, or about 100 times smaller than the outer disk scale lengths and 20 times smaller than the effective radii of the bulge (for NGC~5719, an extreme case, \reff\ is 100 times higher than $h_{2}$).  But they have very high densities.  Extrapolated central surface brightnesses are in the range $11 < \mu_H < 15$, typically 5 mag brighter than the $\mu_0$ of the main galaxy disk.  In many cases, nuclear extended components dominate the surface brightness profile inward of $1\arcsec$ (see Fig.\,\ref{Fig:Profiles}).  

On the basis of their luminosity distribution (Fig.~\ref{Fig:PSvsMK}a) the nuclear extended components might be a heterogeneous family.  Seven galaxies draw a bright sequence with absolute magnitudes $-18 > \MagKExt > -20$, or 3\% to 15\% of the bulge luminosity, while the four remaining cases show fainter luminosities ($-16 > \MagKExt > -17$, or 0.3\% to 3\% of the bulge luminosity) that overlap with the luminosities of the unresolved sources.  These fainter objects are also the smallest ($h_{2} < 10$~pc) and densest, and may be instances of nuclear clusters that we managed to resolve in our images.  The objects in the brighter sequence are also more extended ($25 < h_{2} < 60$~pc), five of the seven cases themselves harbor nuclear unresolved sources, and their sizes and fractional luminosities resemble those of inner disks in elliptical galaxies (Scorza \& Bender 1995; Scorza \& van den Bosch 1998).  
These properties give clues that such objects might have different nature and formation mechanism than their more compact counterparts.   
The presence of more extended components than nuclear clusters indicates that structural deviations from the S\'ersic profile in the nuclei of early-type disk galaxies are not restricted to nuclear star clusters: more extended structures, which we show below to be flattened systems, also cause the bulge profile to deviate from the S\'ersic functional form.  


Scale lengths and central surface brightnesses for inner and outer disks are plotted together in Figure~\ref{Fig:mu0vsh}.  They correlate quite well, both for the nuclear disks alone,

\begin{equation}
I_{0,2}/\LumKSun = 10^{-3.12\pm 0.16} (h_{2}/10\,\mathrm{pc})^{-1.26\pm 0.31},
\label{Eqn:Mu0vsHinner}
\end{equation}
($R_s = 0.83$; \Pnull\ = $8.9\cdot 10^{-3}$), and for inner and outer disks together: 

\begin{equation}
I_{0}/\LumKSun = 10^{-5.30\pm 0.12} (h/\mathrm{kpc})^{-1.07\pm 0.05},
\label{Eqn:Mu0vsHall}
\end{equation}
($R_s = 0.91$; \Pnull\ = $8.4\cdot 10^{-7}$).  In the above equations, $I_{0}$ denotes central intensity in the $K$-band, in units of $K$-band solar luminosities per square arcsecond.  Both relations are consistent with each other, suggesting that $I_{0}\sim h^{-1}$, or $L \sim h$.   Caution is needed when interpreting these relations, given the important selection effects operating on the detection of inner components.  We saw in \S\ref{Sec:ParameterUncertainties} that detection of inner disks is broadly constrained at $-2 <\mu_{0,2} - \mu_{0,\mathrm{Sersic}} < 1$.  In practice, we have found  $\langle\mu_{0,2} - \mu_{0,\mathrm{Sersic}}\rangle = -0.05 \pm 0.81$, i.e., only those inner disks that match the central surface brightness of the bulges's 
S\'ersic profile are detected, and any inner disks much fainter than that limit would go undetected by our  fitting code.  Hence, equations~\ref{Eqn:Mu0vsHinner} and \ref{Eqn:Mu0vsHall} describe the upper envelope of inner disk surface densities.  Similarly, the distribution of our large-scale disks in this diagram is known to define an upper envelope of points (Graham \& de Blok 2001).  

We seek clues on the nature of nuclear extended components by relating their structural parameters to isophotal, dynamical and color information.  Figures with such information are shown for each galaxy in Appendix\,\ref{Sec:Appendix}.  

All of the 11 galaxies with inner extended components show corresponding disky isophotes in the inner arcsec (see Fig.\,\ref{Fig:MuSigColGeom}).  
The association of extended components with disky isophotes is nearly one-to-one, as only three galaxies with disky isophotes in their bulges (NGC~5746, 5965, and 7537) do not require an inner exponential component to fit the surface brightness profile.  NGC~5746, with a peanut-shaped bulge, shows disky isophotes in the bulge-outer disk transition region; NGC~5965, another prototype peanut-shaped bulge, has very strong positive $c_{4}$ disky coefficient in the entire bulge region, suggesting that the entire bulge is disk-like (see, e.g., Bureau et al.~2006); finally, NGC~7537 (Sbc), has a faint bulge and low velocity dispersion ($\sigma$ = 42 \kms).  Hence, the bulges in these three galaxies really look like disks, and fall under the definitions of pseudobulge (Kormendy \& Kennicutt 2004).   For the rest of the sample, inner disky isophotes are associated with nuclear extended components.  

Such correspondence suggests that inner extended components of the surface brightness profiles trace true structural components of the galaxy nuclei, with flattened shapes.  On the basis of SAURON 3D spectroscopy, similar flat kinematically decoupled components aligned with the main galaxy disk have recently been reported for early-type spirals by Falcon-Barroso et al.~(2006; see also McDermid et al.~2006).  
The simple conjecture that they are inner disks, bars, or nuclear rings, is supported by the kinematic data, which show velocities consistent with circular motions and often show velocity dispersion minima.  Smaller, but also extended, 'flattened clusters' have been found in imaging programs for later type spirals by Seth et al.~(2006).   
For our galaxies, the typical outer radius of the positive $c4$ structures is a few hundred pc, which is a typical size of inner bars in double-bar galaxies (Erwin \& Sparke 2002, Erwin 2004).  

For our galaxies, nuclear features in the velocity dispersion profiles (see Fig.\,\ref{Fig:MuSigColGeom}) are not associated with the presence of nuclear extended components.  In the inner arcsec where typically nuclear components have a strong contribution to the total surface brightness, most velocity dispersion profiles are quite flat, with specific instances of central peaks or drops.  These profiles, from minor-axis spectra taken with the WHT/ISIS spectrograph, have a 0.3 arcsec sampling and a seeing of around 1.0 arcsec (Falc\'on-Barroso et al.\ 2003).  Clearly, two-dimensional spectroscopy at sub-arcsecond resolution is required for a kinematic characterization of the nuclear extended components.  

The colors of the extended nuclear components are quite red, and, as shown in Paper~I, they correspond to a stellar population reddened with a mean of $A_V = 0.5-1.0$.  The color profiles, however, are generally very smooth  (see Fig.\,\ref{Fig:MuSigColGeom}).  
This suggests that the stellar populations of the nuclear extended components are not very distinct from those of the surrounding bulges.  In particular, despite signs of nuclear star formation (Paper~I), the bulk of their stellar populations are not recent additions to the galaxy.  Note that the combination of blue and NIR colors is particularly sensitive to population age.  In later-type spirals, the nuclear 'flattened clusters' are bluer than the host spheroid (Seth et al.~2006), perhaps indicating an extended star formation history for the nuclear structures which continues today for late-type spirals (see also Walcher et al.~2006), but has finished in most earlier type disk galaxies.  


\begin{figure}[htbp]
\begin{center}
\includegraphics[width=0.5\textwidth]{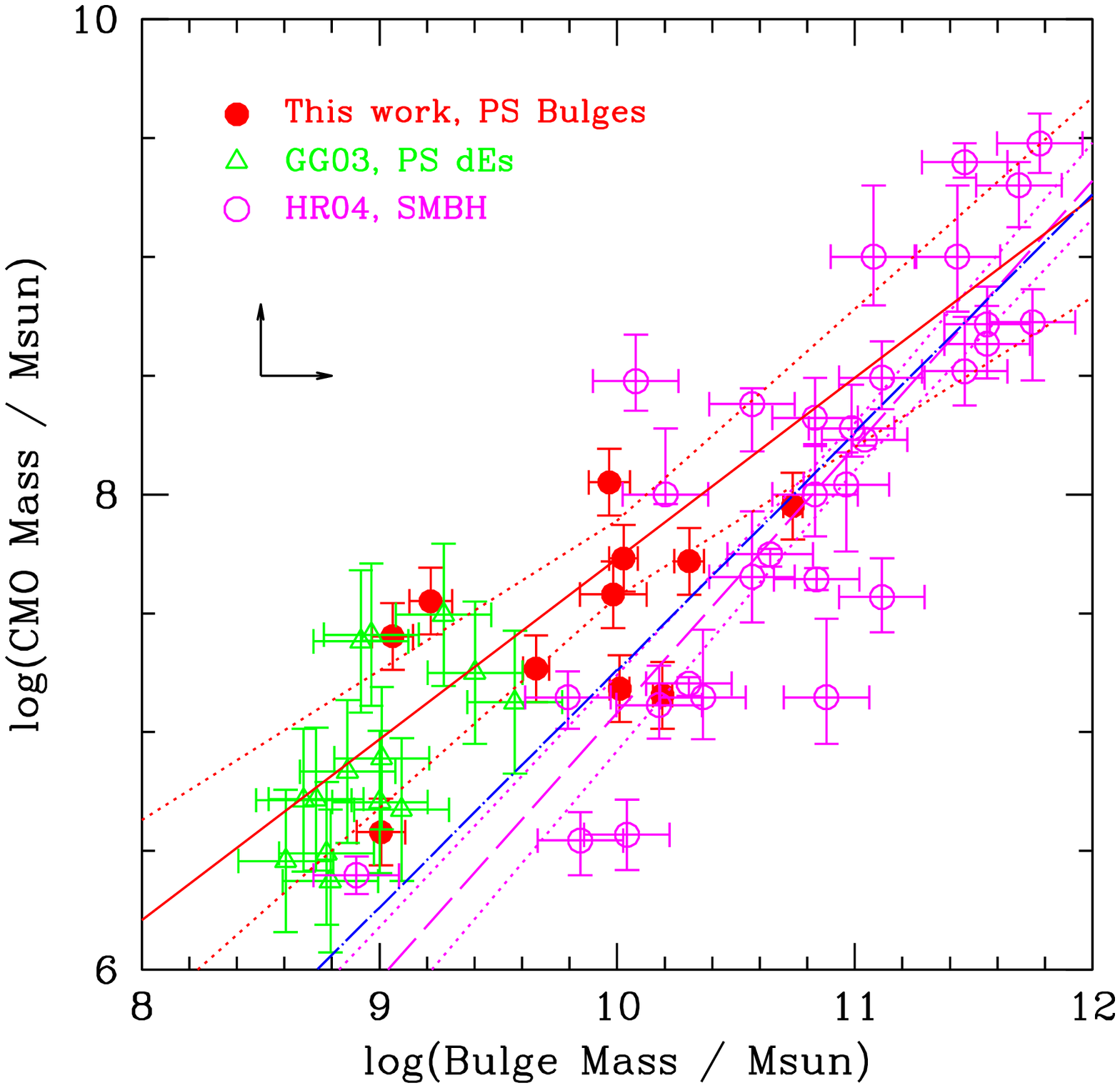}
\caption{The spheroid mass is plotted against the mass of the compact massive object.  
\textit{Filled circles: } this work, bulge mass vs nuclear unresolved source mass. 
\textit{Triangles: } Coma dwarf ellipticals from Graham \& Guzman (2003).  The above are photometric masses, in Solar units, derived from the $K$-band luminosities assuming $M/L_{K}=0.8$, see \S~\ref{Sec:PointSources}.  
\textit{Open circles: } ellipticals and bulges with central SMBHs, from H\"aring \& Rix (2004).  
\textit{Solid line: } regression for bulges and dEs (Eqn.~\ref{Eqn:MassNuc}).
\textit{long-dash line: } bisector linear regression for the sample of H\"aring \& Rix (2004), see equation~\ref{Eqn:SMBHmassHR04}.
\textit{dotted lines: } $\pm 1\sigma$ range of acceptable fits. 
\textit{Dot-dashed line: } the relation between CMOs and galaxy mass derived by F06 under the assumption of a linear scaling (Eqn.~\ref{Eqn:MassCMOF06}).  
}
\label{Fig:CMOvsBul}
\end{center}
\end{figure}

\subsection{Nuclear unresolved sources}
\label{Sec:PointSources}

As previously noted, we detect unresolved sources (referred to as point sources, or PSs) in 11 of our 19 galaxies; five of these coexist with an inner exponential.  Our detection frequency is similar to that reported in previous studies: Ravindranath et al.\ (2001) report a 50\% detection fraction; B\"oker et al.\ (2002) find nuclear PSs in 76\% of their sample; C\^ot\'e et al.~(2006) report a fraction of 66\% to 88\% of nucleated early-type galaxies from the ACS Virgo Cluster Survey.  

For our sample, the absence of AGN nuclei, and the nuclear colors suggest that light from the PSs is stellar in origin, i.e.\ that they are nuclear star clusters.  A similar conclusion is reached by Phillips et al.\ (1996) and by Carollo et al.\ (1998), whereas Ravindranath et al.\ (2001) argue for a non-stellar origin for the PS light on the basis of the high frequency of AGN in their sample.  In what follows we assume the PSs to be nuclear clusters, although we recognize that, in general, AGN contribution to nuclear emission may be important in specific samples.  

Absolute magnitudes are in the range $-13 > \MagKPt > -17$, (Table~\ref{Tab:NucPhysParameters}), on the mean 6 mag fainter than their host bulges ($K$-band luminosity ratio 0.4\%).  Their luminosities correspond to 10-20 globular clusters, quite comparable to nuclei in nucleated Virgo ellipticals (C\^ot\'e et al.~2006).   These absolute magnitudes do not vary systematically with galaxy inclination (Fig.\,\ref{Fig:PSvsMK}d), suggesting that extinction in the parent disks do not affect \MagKPt. Extinction at the nuclei themselves does have a small effect on \MagKPt.  We show for reference an $A_{V}=1$ mag extinction vector in Figure\,\ref{Fig:PSvsMK}c.  



We showed in Paper II that \MagKPt\ correlate with the bulge absolute magnitude, a result later confirmed by Rossa et al.\ (2006) using optical imaging.  The correlation given in Paper~II, eqn.~1, was derived from fits to 17 objects, using a slightly wider PSF (0.19\arcsec\ FWHM) than used here (0.131\arcsec).  For the fits presented here, an  orthogonal regression to the \MagKPt--\MagKBul\ distribution gives

\begin{equation}
\LumKPt/\LumKSun = 10^{7.70\pm 0.17} (\LumKBul/10^{10}\,\LumKSun)^{0.63\pm 0.37},
\label{Eqn:PSvsMK}
\end{equation}
\noindent (using $\MagKSun = 3.41$, Allen 1973), 
consistent with that we gave in Paper II.  However, with only 11 data points and a significant scatter, the relation is not statistically significant for this sample ($R_s = 0.43$; \Pnull\ = $0.18$). 
Graham \& Guzm\'an (2003), who analyzed Coma dwarf elliptical (dE) $HST$/WFPC2 $F606W$ galaxy surface brightness profiles using the same fitting code as the present paper, also find a scaling between PS and bulge luminosity, with a consistent slope of $0.87 \pm 0.26$. Their data points are given in Figure\,\ref{Fig:PSvsMK}a\footnote{After applying a constant color term $F606W - K = 2.7$.  This scaling corresponds to an old population with 0.4 times solar metallicity, using the models of Vazdekis et al.\  (1996); a lower metallicity would make the total and nuclear luminosities of the dE fainter.}.  
Inasmuch as the nuclei of our bulges and those of the Coma dEs are similar structures, we may use the combined sample to derive the scaling of nuclear unresolved sources with spheroid luminosity.  We find

\begin{equation}
\LumKPt /\LumKSun = 10^{7.75\pm 0.15} (\LumKBul/10^{10}\,\LumKSun)^{0.76\pm 0.13}.
\label{Eqn:PSdEvsMK}
\end{equation}
($R_s = 0.72$; \Pnull\ = $4.0\cdot 10^{-4}$).  We transform eqn.~\ref{Eqn:PSdEvsMK} into a mass relation, i.e., 

\begin{equation}
\MassPt/\MassSun = 10^{7.73\pm 0.16} (\MassBul/10^{10}\,\MassSun)^{0.76\pm 0.13}, 
\label{Eqn:MassNuc}
\end{equation}
\noindent where $M/L_{K} = 0.8$ has been assumed for nuclei and bulges, from Bell \& de Jong 2001, for the typical colors of bulge populations.  The normalization is insensitive to the choice of $M/L_{K}$ because its slope is close to unity;  using the more extreme $M/L_{K} = 0.5$ would yield a nearly identical normalization term of $10^{7.68 \pm 0.16}$.  

Equation~\ref{Eqn:MassNuc} may be compared to scaling regressions found in other studies of  nuclear clusters.  F06 find $\MassNuc \sim \MassBul^{1.32 \pm 0.25}$, which is $\sim2\sigma$ from our slope.  And, for their so-called compact massive objects (CMOs), a class encompassing central supermassive black holes (SMBHs) and central star clusters, F06 find, when fixing the exponent to unity,

\begin{equation}
\MassCMO/\MassSun = 10^{7.26\pm 0.47} (\MassGal/10^{10}\,\MassSun)^{1.0}. 
\label{Eqn:MassCMOF06}
\end{equation}

\noindent while, for a sample of galaxies with kinematic SMBH mass determinations, H\"aring \& Rix (2004, hereafter HR04) infer

\begin{equation}
\MassSMBH/\MassSun = 10^{7.08\pm 0.10} (\MassBul/10^{10}\,\MassSun)^{1.12\pm 0.06}. 
\label{Eqn:SMBHmassHR04}
\end{equation}

Figure~\ref{Fig:CMOvsBul} shows the host spheroid mass against the nuclear mass of unresolved bulge and dE components.  For comparison we plot the HR04 distribution of \MassSMBH\ vs host mass.  The regressions provided by HR04 (eqn.~\ref{Eqn:SMBHmassHR04}), F06 (eqn.~\ref{Eqn:MassCMOF06}), and us (eqn.~\ref{Eqn:MassNuc}) are plotted as well.  
The distribution of \MassPt\ converges with that of the \MassSMBH\ regression at the high-mass end, but it progressively deviates as we move to lower spheroid masses.  Both our distribution of \MassPt\ and HR04's distribution of \MassSMBH\ show a large scatter at $\log(\MassBul) \leq 10$, hence the slopes of the regressions have large uncertainties in that mass domain.  
However the nuclear cluster masses are systematically above the extrapolation of the HR04 or F06 relations.  From equation~\ref{Eqn:MassNuc}, nuclei fractional mass ($\MassPt / \MassBul$) increases from 0.19\%, identical to central black holes fractional masses, to 0.71\%, as \MassBul\ decreases from $10^{11.5}$\MassSun\ to $10^{9.5}$\MassSun.  We are aware that different mass determinations are being compared in Figure~\ref{Fig:CMOvsBul} (photometric masses for our data, virial masses for F06, and masses derived from Jeans equations' modeling for HR04).  Nevertheless, we argue that the observed offset between our unresolved components and the SMBH relations from either HR04 or F06, cannot be explained by uncertainties in our mass determinations.  Bridging the offset by changing our assumed $M/L_{K}$ would require $M/L_{K}$ of nuclei to be lower than those of their parent galaxies by a factor of $\sim$4, which is implausible for stellar populations without signs of vigorous star forming activity. Using dynamical masses for our bulges, instead of the photometric masses, does not help either: applying for consistency the same form of the Virial mass used by F06,

\begin{equation}
\MassBulDyn \equiv 5\,\reff\,\sigma^{2}/G
\label{Eqn:VirialMass}
\end{equation}

\noindent the offset actually increases (\MassBulDyn\ are on the mean a factor 2 \textsl{below} the stellar-mass \MassBul) which shifts our points to the left, away from the F06 relation; such discrepancy indicates that eqn.~\ref{Eqn:VirialMass} underestimates the true dynamical mass of the bulges -- most likely due to ignoring the rotational kinetic energy.  

We thus cannot reproduce F06's result that nuclear star clusters masses fall onto the same linear relation as defined by more massive central black holes.  Any CMO--bulge mass relation that encompasses both central black holes and nuclear star clusters must be non-linear for bulge masses of say $\MassBul \leq 10^{10}$\MassSun.  An orthogonal regression to the nuclear star cluster masses and the black hole masses, against the host bulge mass, gives

\begin{equation}
\MassCMO/\MassSun = 10^{7.51\pm 0.06} (\MassBul/10^{10}\,\MassSun)^{0.84 \pm 0.06}. 
\label{Eqn:MassCMOus}
\end{equation}

We conjecture that F06 find a linear \MassCMO --\MassBul\ relation because, near the low mass end, their sample includes many S0's.  When deriving Virial masses using equation~\ref{Eqn:VirialMass}, because their \reff\ derive from single S\'ersic fits to the galaxy profiles (C\^ote et al. 2006), spheroid masses are probably overestimated for all of their S0 galaxies, which is roughly half of their sample. Laurikainen et al.\ (2005) report that the mean bulge-to-total ratio for lenticular galaxies is 0.24, so one may expect a downward correction to (half of) the F06 spheroidal masses of $\sim$0.6 dex, which would explain the different trends in F06 and us.  


\begin{figure}[htbp]
\begin{center}
\includegraphics[width=0.48\textwidth]{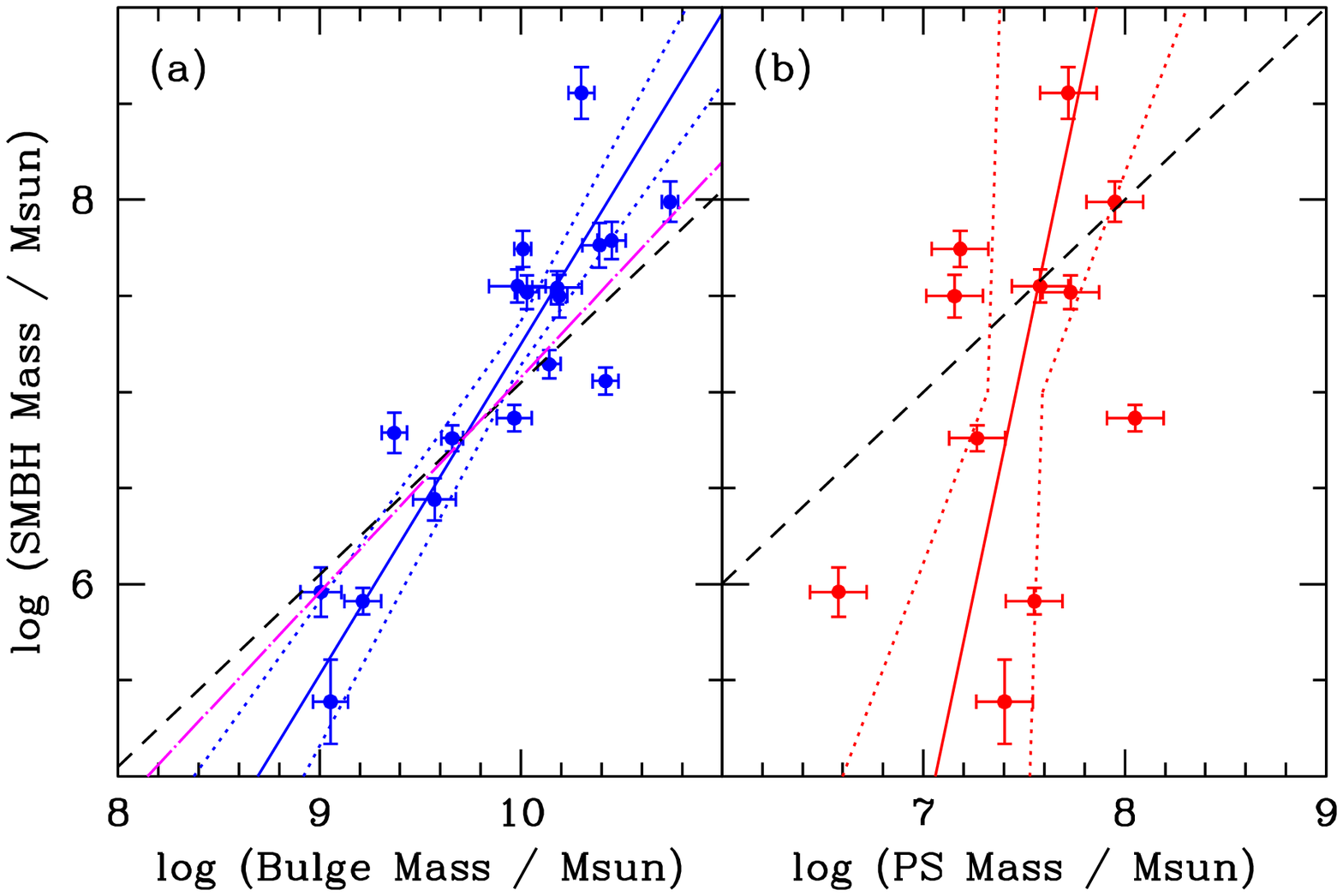}
\caption{Central supermassive black hole mass, estimated from the velocity dispersion following Tremaine et al.~(2002), plotted vs.\ (\textit{a}) the photometric bulge mass, and 
(\textit{b}) the photometric mass of the nuclear unresolved sources.  
Photometric masses are derived from the $K$-band luminosities assuming $M/L_{K}=0.8$, see \S~\ref{Sec:PointSources}.  
The \textit{solid lines} in panels (\textit{a}) and (\textit{b}) correspond to eqns.~\ref{Eqn:MassBulSMBH} and \ref{Eqn:MassPtSMBH}, respectively, while \textit{dotted lines} delineate the approximate region of acceptable fits given the 1-$\sigma$ error bars in the fit coefficients.  
In panel (\textit{a}), the \textit{dot-dashed line} traces the SMBH mass -- shperoid mass relation of HR04, while the \textit{dashed line} traces a slope unity relation for reference.    
In panel (\textit{b}), the \textit{dashed line} is the locus where PS masses and SMBH masses are equal. }
\label{Fig:SMBHmass}
\end{center}
\end{figure}

\subsection{Nuclear black hole masses}
\label{Sec:SMBH}

To further investigate the connection between star clusters and central SMBHs proposed by F06 and Wehner \& Harris (2006), in this subsection we explore the scalings of SMBH mass \MassSMBH, nuclear cluster mass, and bulge mass.  We estimate \MassSMBH\  using the expression 

\begin{equation}
\log \MassSMBH = 4.02 (\pm 0.32) \log (\sigma/200) + 8.13(\pm 0.06)
\label{Eqn:BHsigmaTR02}
\end{equation}

\noindent from Tremaine et al.\  (2002 hereafter T02), where \MassSMBH\ is in Solar masses, and $\sigma$ is the central velocity dispersion in \kms.  
We show bulge masses against SMBH masses in Figure~\ref{Fig:SMBHmass}a.  An orthogonal regression of our estimated \MassSMBH\ against \MassBul\ gives 

\begin{equation}
\MassSMBH/\MassSun = 10^{7.25\pm 0.11} (\MassBul/10^{10}\,\MassSun)^{1.72 \pm 0.26}, 
\label{Eqn:MassBulSMBH}
\end{equation}
($R_\mathrm{S} = 0.84$; \Pnull\ = $5.6\cdot 10^{-4}$).  Such a strong relation is expected because it is largely a manifestation of the strong Faber-Jackson relation followed by our bulges.  Our bulges follow $\LumKBul \propto \sigma^{2.86 \pm 0.5}$ (see Paper~IV, \S~3.1.5.), and using $\MassSMBH \propto \sigma^{4.02 \pm 0.32}$ we expect $\MassSMBH \propto (\MassBul)^{4.02 / 2.86} = (\MassBul)^{1.41 \pm 0.31}$, consistent with eqn.~\ref{Eqn:MassBulSMBH}.  
Equation~\ref{Eqn:MassBulSMBH} is also consistent with that found by Laor (2001) for a sample of active and inactive galaxies, i.e., $\MassSMBH \propto (\MassBul)^{1.53 \pm 0.14}$. Our scaling lies 2$\sigma$ from the relation found by HR04 (drawn in Figure~\ref{Fig:SMBHmass}a), and formally departs 2.8$\sigma$ from linearity (shown with a dashed line in Figure~\ref{Fig:SMBHmass}a). We do not want to give much weight to this departure from linearity, given the small sample size, and the fact that we are extrapolating the $\MassSMBH - \sigma$ relation faintward.   

If central black holes and central star clusters are intimately related, the latter perhaps being failed black holes, then the $\MassSMBH - \sigma$ relation (eqn.~\ref{Eqn:BHsigmaTR02}) should be able to predict the nuclear cluster masses.  We plot \MassPt\ (measured) against \MassSMBH\ (estimated) in Figure~\ref{Fig:SMBHmass}b, together with the result of an orthogonal regression to the nuclear star-cluster--black hole mass distribution, which gives
\begin{equation}
\MassPt/\MassSun = 10^{7.46\pm 0.14} (\MassSMBH/10^{7}\,\MassSun)^{0.20 \pm 0.16}, 
\label{Eqn:MassPtSMBH}
\end{equation}
($R_s = 0.37$; \Pnull\ = $0.24$).  The correlation is not statistically significant, partly due to the small sample size and the scatter of the data points.  We report this regression to clarify that the slope of the relation is several sigma away from the value of 1 expected if the \MassSMBH--$\sigma$ relation had predicted the masses of all the star clusters.  This occurs also when SMBH masses are computed following Merritt \& Ferrarese (2001).  

Wehner \& Harris (2006) advocate a CMO transitional mass at $10^7 \MassSun$.  While our \MassSMBH\ scatter above and below this mass limit, all but one of our nuclear star clusters lie above it. 
Some bulges are already known to contain both a SMBH and a nuclear star cluster: NGC~7457, which is in our sample, and NGC~3384 (Wehner \& Harris 2006).  
Curiously, above the $10^7 \MassSun$ limit our nuclear star cluster masses agree with the predicted SMBH masses.  Below the $10^7 \MassSun$ limit, cluster masses lie $\sim$1 dex above the extrapolation of the black-hole mass relation (shown as a dashed line in Figure~\ref{Fig:SMBHmass}b).  Hence, again, the \MassSMBH--$\sigma$ relation appears not to predict the nuclear star cluster masses below this limit.  It is unlikely that our surface brightness profile fitting code has overestimated the PS light by such large factors (\S\,\ref{Sec:ParameterUncertainties}), or that the $M/L$ we adopted for the nuclear clusters is off by similar amounts.  Some of the deviation may arise from the large intrinsic scatter in the $\MassSMBH - \sigma$ distribution in the low-mass range $6 < \log(\MassSMBH) < 8$; see, e.g., Figure~7 from T02.  

\begin{figure}[htbp]
\begin{center}
\includegraphics[width=0.45\textwidth]{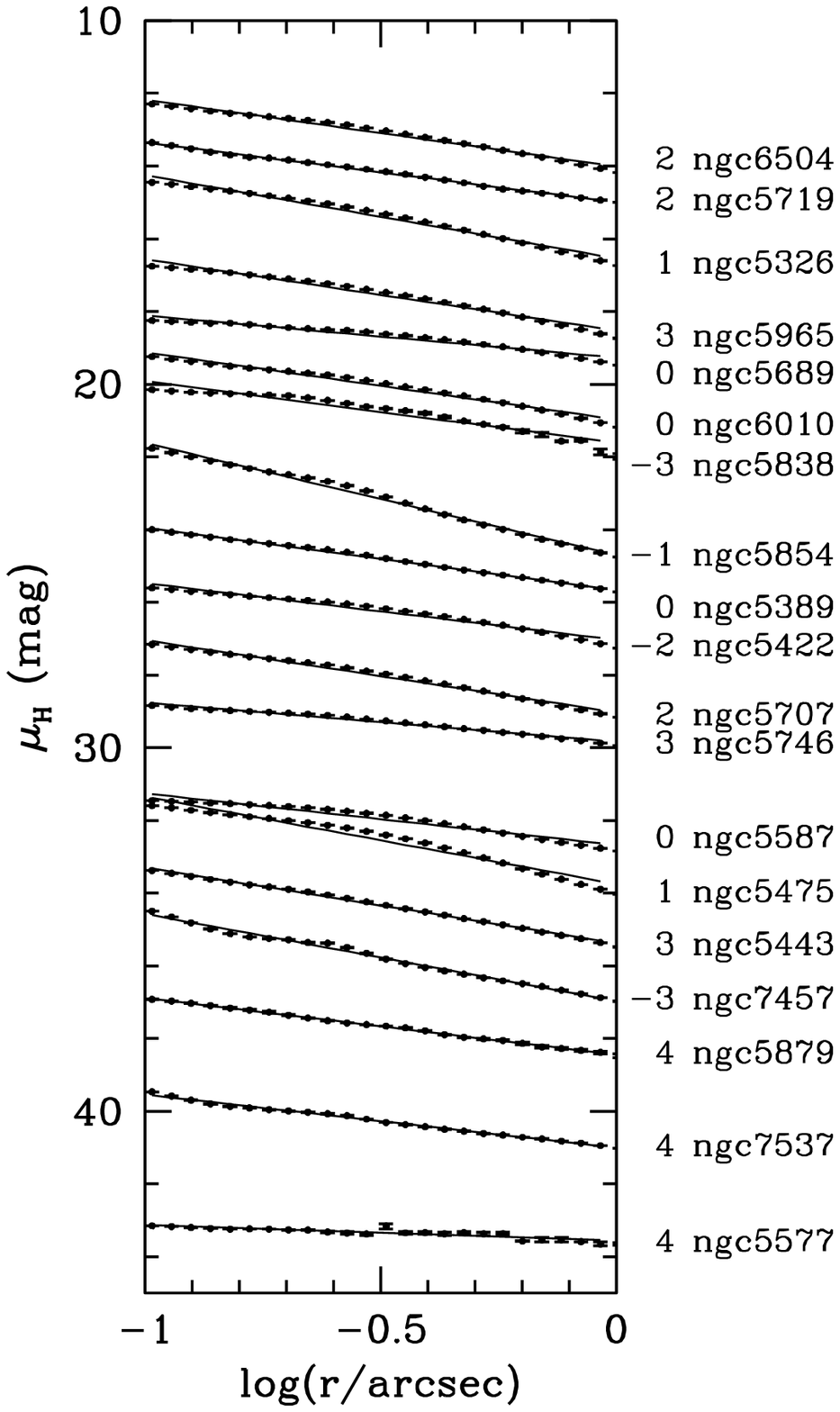}
\caption{Inner surface brightness profiles plotted against $\log(R)$ in the range $0.1 \leq R \leq 1$ arcsec ($R\equiv$ geometric mean radius).  Profiles are sorted by decreasing bulge absolute magnitude, brightest above, and are offset by 1.5 mag/arcsec$^2$ from each other for clarity.  {\sl Solid lines:} direct least-squares power-law fits, with exponent $\gamma$ given in Table\,\ref{Tab:Gammas}, col.\ 2.   }
\label{Fig:InnerProfiles}
\end{center}
\end{figure}

\begin{figure}[htbp]
\begin{center}
\includegraphics[width=0.5\textwidth,angle=-90]{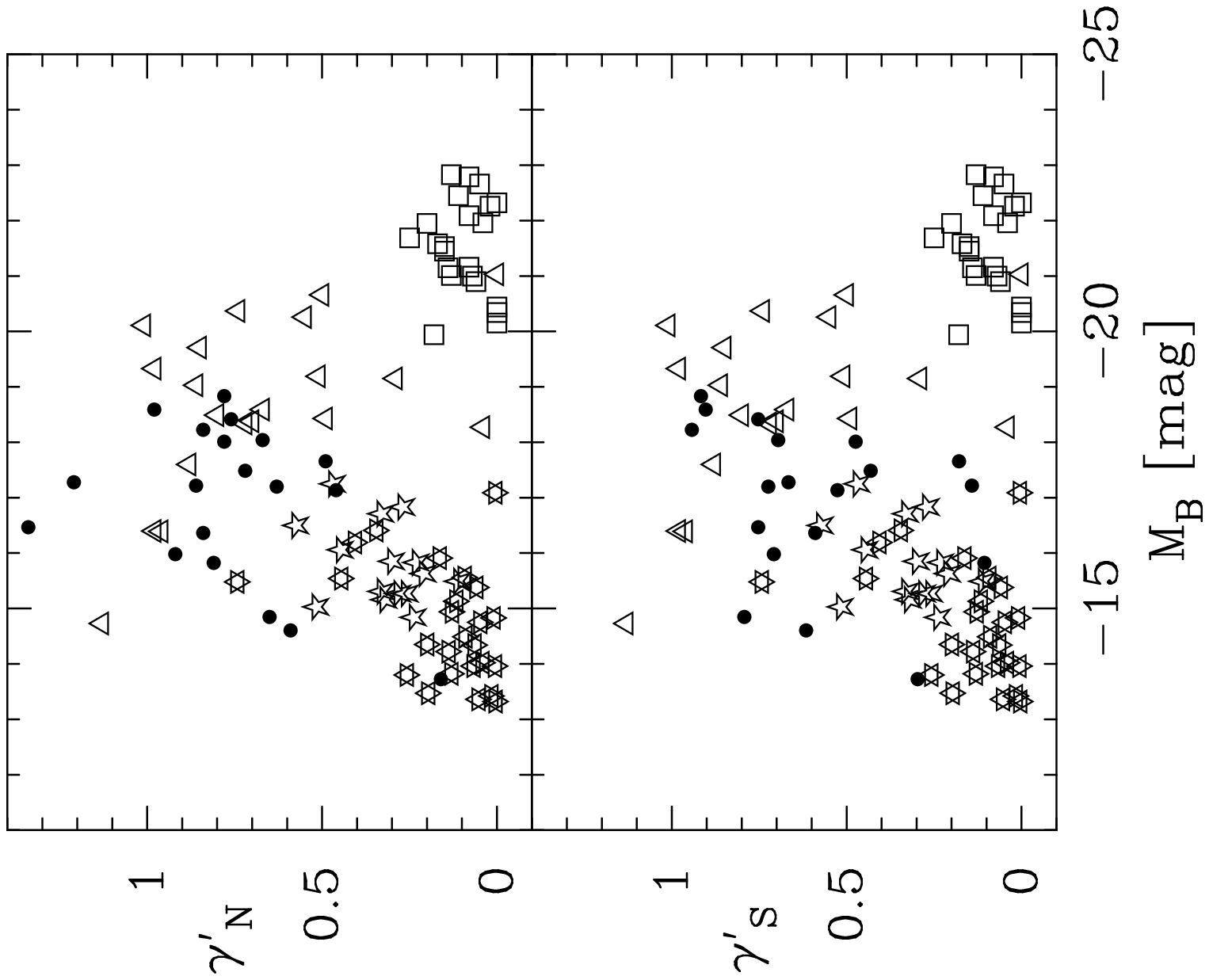}
\caption{The spheroid $B$-band absolute magnitude plotted against two inner profile slope determinations $\gamma$. 
\textit{Top panel:} slope $\gammaNukPrime$ of the best-fit Nuker model at 0.3 arcsec.  
\textit{Bottom panel:} slope \gammaSerPrime\ of best-fit S\'ersic profile at $0.15\,\reff$. 
{\it Filled circles}: bulges, this work.
{\it Squares}: core ellipticals from L95. 
{\it Triangles}: power-law ellipticals from L95.
{\it Five-pointed stars}: dwarf ellipticals from Graham \& Guzm\'an (2003).
{\it Six-pointed stars}: dwarf ellipticals from Stiavelli et al.\ (2001).
}
\label{Fig:GammaMB}
\end{center}
\end{figure}

\input tabgammas.tex

\subsection{Inner profile slopes}
\label{Sec:Gammas}

The {\it HST}-resolved, logarithmic slopes of inner surface brightness profiles are a useful galaxy parameter as they can test the applicability of density profile models, which are often based on power-laws (e.g., Jaffe 1983; Hernquist 1990; Dehnen 1993; Tremaine et al.\ 1994). Inner slopes constrain the shape of the potential and the types of orbits that may be present in the nuclei.  On the observational side, interest in profile slopes arose from the discovery of a bimodal distribution of inner slopes in samples of intermediate- and high-luminosity elliptical galaxies (Ferrarese et al.\ 1994; L95; Gebhardt et al.\ 1996) which suggested different formation/evolution processes for the cores of giant and less luminous ellipticals, perhaps linked to the presence of binary black holes (e.g., F97).  

For our galaxies, all surface brightness profiles continue to rise inward to the resolution limit of the data (Fig.\,\ref{Fig:InnerProfiles}).   Inner profiles approximate power laws, although small but clear deviations are evident in Figure\,\ref{Fig:InnerProfiles}, making the slope determination dependent on the radial range used for the measurement.  The presence of inner components further complicates matters and means that two types of profile slopes may be measured: the slope of the {\it total} surface brightness profile, or that of the underlying bulge component.  Both have their own merits.  The latter provides a cleaner measure of the slope of the spheroidal component of the bulge, unaffected by inner disks/bars or point sources.  To date, common practice has been to avoid such nuclear features when deriving the inner profile slope.  On the other hand, a direct measure of the slope has the advantage of being independent of any model decomposition.
Moreover, if the central supermassive black holes, known to reside in ellipticals and bulges (Kormendy \& Richstone 1995) have grown adiabatically within the nuclear star clusters (e.g., Young 1980; van der Marel 1999), then the profile slope of the central cluster is of interest and should not be avoided --- although resolution does become an issue.  

Several methods have been employed in the past to measure the inner profile slope denoted by $\gamma$: a single power-law (e.g., Phillips et al.\ 1996); a double power-law (Ferrarese et al.\ 1994); a Nuker fit (L95); a measure of the logarithmic derivative of the best-fit Nuker-law at either some fixed radius (usually denoted as $\gamma\prime$, e.g., R01, and here denoted \gammaNukPrime) or over some small interval, e.g.\ 0.1-0.5 arcseconds (usually denoted as \gammaNukAverage; e.g.\ Stiavelli et al.\ 2001); and the logarithmic derivative of the best-fit S\'ersic model (denoted here as $\gammaSerPrime$; Graham \& Guzm\'an 2003).  For the S\'ersic model, \gammaSerPrime\ can easily be evaluated at any radius $R$, such that 

\begin{equation}
\gammaSerPrime(R) \equiv -d\log I(R)/d\log R =\frac{b_n}{n}\left(\frac{R}{\reff}\right)^{1/n},
\label{Eqn:GammaSersic}
\end{equation}

\noindent where $b_n \approx 1.9992 n - 0.3271$, and \reff\ is the effective radius.  

For our sample, we have computed the inner profile slope following all of the methods outlined above,  except for \gammaNukAverage, for lack of a precise definition of this quantity.  
The various determinations of the inner profile slopes are given in Table~\ref{Tab:Gammas}.  
These include direct power-law least-squares fits to the profiles over ranges of $0.1< R/{\rm arcsec} < 1$ and $20 < R/{\rm pc} < 200$; $\gammaNukPrime(0.3\arcsec)$; $\gammaSerPrime(0.3\arcsec)$; and, \gammaSerPrime(0.15\,\reff).  Comparison of the results indicates that direct power-law fits over ranges of $0.1< R < 1$ arcsec or $20 < R/{\rm pc} < 200$ yield slopes $\gamma$ quite similar to $\gammaNukPrime(0.3\arcsec)$ derived from a Nuker-fit.  Our Nuker fits extended inward to $R=0.1\arcsec$, hence the  similarity of $\gamma\prime(0.3\arcsec)$ to a direct power-law fit is not surprising: these three fits encompass any compact nuclear components.  The values of $\gammaSerPrime(0.3\arcsec)$ and \gammaSerPrime(0.15\,\reff), which come from the S\'ersic fits to the host bulge (\S~\ref{Sec:BulgeDiskFits}), are equal or lower than the other values.  They provide the slope corresponding to the underlying bulge components, and, as such, they are the most directly comparable to the \gammaNukAverage\ values presented by e.g., Carollo \& Stiavelli for late-type spirals, and those of Stiavelli (2001) for dwarf ellipticals.  

Figure\,\ref{Fig:GammaMB} shows the $B$-band absolute magnitudes of the bulges
plotted against the values of \gammaNukPrime\ derived at 0.3\arcsec, and of \gammaSerPrime\ derived 
at $R/\reff =0.15$. 
The $B$-band absolute magnitudes of the bulges were derived from the galaxy total corrected $B$-band magnitudes from the  RC3 and the bulge-to-disk ratios derived in this paper; this procedure avoids $B$-band bulge-disk decompositions which are heavily affected by dust. 
Also included in the figure are the elliptical galaxy data from L95
(Nuker-model $\gamma$)
and the dwarf elliptical data from Stiavelli et al.\ (2001; mean slope \gammaNukAverage\  between 0.1$\arcsec$ and 0.5$\arcsec$) and Graham \& Guzm\'an (2003; \gammaSerPrime\ at $R=0.2\arcsec$ from S\'ersic fit).  
Inner slopes for bulges cluster above $\gamma = 0.5$ when the light of nuclear components is included (top panel of Fig.\,\ref{Fig:GammaMB}), and matches that of ellipticals of similar luminosities.  When nuclear components are excluded (bottom panel of Fig.\,\ref{Fig:GammaMB} ), slopes trace a continuous distribution in the range $0 < \gamma < 1$. 

In a previous study of inner profile slopes of galaxy bulges, Carollo \& Stiavelli (1998) obtain a similar range of $0<\gamma<1$ values to us.  They however differ in the interpretation of the results.  Because their distribution of \gammaNukAverage\  shows two distinct clumps, they argue that two separate families of bulges exist, i.e.\ "exponential" ($\gammaNukAverage \sim 0.3$) and "\r14" ($\gammaNukAverage \sim 0.8$) bulges.  Figure\,\ref{Fig:GammaMB} shows that our bulges do not cluster in two clumps, instead they cover a continuous range from the low to the high values of the Carollo \& Stiavelli distribution.  The origin for their bi-modal distribution of \gammaNukAverage\  is unclear.  Their \gammaNukAverage\ values come from Nuker fits and not from exponential or \r14\ fits, hence the bi-modality should not be a consequence of their splitting the sample into "exponential" and "\r14" classes.  
But two aspects of their analysis could lead to a polarization of their \gammaNukAverage\ toward high and low values.  
One of these is the subjective choice of the radial range for the Nuker fits.  
While nuclear components are easy to identify by eye when the underlying profile is shallow, such as in an exponential bulge, for galaxies with steeper underlying bulge profiles, ($n\ga 2$), the 'break' from the bulge to the nuclear component becomes weaker and easier to miss by eye, hence the radial range occupied by the nuclear component can easily be included when performing the Nuker fit, yielding a biased,  higher value of \gammaNukAverage\ than that corresponding to the underlying bulge. Simultaneous fits to the bulge and the nuclear component are needed in those cases to  remove such bias from the determination of the bulge profile slope.  
Another aspect that may polarize the \gammaNukAverage\ distribution into high and low clumps may be the derivation of bulge absolute magnitudes without a bulge-disk decomposition, and appying exponential- or \r14-constrained fits to profiles from small-field HST/WFPC2 images: \r14\ models are known to over-estimate the flux of \rn\  systems if $n<4$, while, exponential models are known to underestimate the flux in \rn\ systems which have $n>1$ (e.g., Graham 2001a).
Finally, sample selection may also be important.  Their sample includes many barred galaxies, and is overall later type than ours, though we note that many of our bulges show exponential-like profiles, i.e., profiles which Carollo \& Stiavelli associate with the low-\gammaNukAverage\  clump.  
The above arguments, together with the distributions shown in Figure~\ref{Fig:GammaMB}, suggest that the slopes of the total galaxy profiles (including nuclear components), cluster around $0.5 < \gamma < 1$, but \textit{bulge profiles as a class cover a continuous distribution of nuclear slopes in the range $0 < \gamma < 1$}.  

\section{Conclusions}
\label{Sec:Conclusions}

At \HST\ resolution, nuclear photometric components, in addition to the S\'ersic bulge and the exponential outer disk, are exceedingly common ($\sim$90\%) in early- to intermediate-type disk galaxies.  
Spatially-resolved nuclear components are found in 58\% of our sample.  These components are geometrically flat systems, and could be disks, bars or rings.  The ones detected have comparable central surface brightness to the underlying bulges, but fainter such systems may exist.  The isophotal signatures indicate total sizes of a few hundred pc, similar to those of inner bars in double-barred galaxies.  Often, such components are reddened by dust; the evidence from optical and NIR colors, presented in Paper~I, as well as their high densities, suggest that they are old rather than late additions to the bulges.  




A majority of the galaxies ($\sim$58\%) harbor sources unresolved by \HST/NICMOS2.  They are most likely star clusters, with luminosities corresponding to 10--20 globular clusters, and similar to other unresolved sources found in the nuclei of ellipticals, dwarf ellipticals and bulges.   When combined with similar nuclear components in dE galaxies, their photometric masses scale with spheroid mass as 
$\MassPt/\MassSun = 10^{7.73\pm 0.16} (\MassBul/10^{10}\,\MassSun)^{0.76\pm 0.13}$.  
Our central star clusters fall above the faint-ward extrapolation of \MassSMBH--\MassBul\ relations derived by HR04 or F06.  In order to extend a CMO-style relation to faint spheroid luminosities, a moderate non-linearity is needed, and we propose the relation  
$\MassCMO/\MassSun = 10^{7.51\pm 0.06} (\MassBul/10^{10}\,\MassSun)^{0.84 \pm 0.06}$.  
But we see additional difficulties with the CMO picture in that  
all of our PS show masses above the cluster-black hole transitional mass of $10^{7} \MassSun$ proposed by Wehner \& Harris (2006).   

Bulge surface brightness profiles rise inward to the limit of the \HST/NICMOS resolution, $\sim$10 pc for the current sample.  While the inner bulge profiles deviate from pure power-laws, "break radii" in a Nuker-law sense are not present.  Structurally, the bulges of early- to intermediate-type galaxies may be globally grouped with the "power-law", intermediate- and low-luminosity elliptical galaxies.  
Negative logarithmic nuclear profile slopes of the S\'ersic bulge components  
$\gamma$ cover a continuous range of $0<\gamma<1$, 
overlapping with dwarf-ellipticals at the faint end, and with intermediate luminosity ellipticals at the bright end.  We find no evidence to support a bimodal distribution of $\gamma$ reported by others.  


\acknowledgements This research has made use of the NASA/IPAC Extragalactic Database (NED) which is operated by the Jet Propulsion Laboratory, California Institute of Technology, under contract with the National Aeronautics and Space Administration.  This research has made use of the HyperLeda database.  The United Kingdom Infrared Telescope is operated by the Joint Astronomy Centre on behalf of the U.K. Particle Physics and Astronomy Research Council.

\appendix

\section{Surface brightness, isophotal, dynamical, and color profiles}
\label{Sec:Appendix}

\setcounter{figure}{0}

Figure\,\ref{Fig:MuSigColGeom} shows the surface brightness profiles together with dynamical, isophotal and color profiles for each galaxy.   The top panel shows the combined HST+GB $\mu_H$ profile, and S\'ersic+exponential fit carried out excluding the inner 1 arcsecond.  Beneath the galaxy names is the fit type code corresponding to the full fit including additional nuclear components, from Table~\ref{Tab:FitParameters}.  The second panel gives the residuals from the fit.  
The third panel shows the minor-axis velocity dispersion profiles, folded around the origin, from Falc\'on-Barroso et al.\ (2003); the abscissa has been scaled to the geometric mean radius using the mean of the ellipticity profile.  The fourth and fifth panels show the combined HST+GB minor-axis $B-I$ and $I-H$ color profiles on the side of the bulge not seen through the disk (the "dust-free" side), from Paper\,I; the abscissa has been scaled to the geometric mean radius using the mean of the ellipticity profile.  The sixth panel shows the ellipticity profile, and the bottom panel shows the 4th order Fourier cosine term $c4$, from the ellipse fits; inside 6 arcsec ({\it triangles}) the latter two profiles are derived from HST/NICMOS F160W images, while outside 6 arcsec ({\it squares}) they are derived from the UKIRT $K$-band images (Peletier \& Balcells 1997).  

\begin{figure*}[p]
\begin{center}
\caption{Profiles of surface brightness, velocity dispersion, $B-I$, $I-H$, ellipticity, and $c4$ diskiness/boxiness coefficient.  Abscissa is the logarithm of the geometric mean axis in arcsec.  $\mu_H$: combined HST+GB surface brightness profile, and model fits excluding the central arcsec ({\it dashed line}: S\'ersic bulge; {\it dotted line}: exponential disk).  $\Delta\mu_H$: residuals from the fit (data {\it minus} model).  $\sigma$: minor-axis velocity dispersion profiles from Falc\'on-Barroso et al.\ (2003), folded around the origin ({\it triangles}: dust-free side; {\it inverted triangles}: side seen through the disk; abscissa has been projected to the geometric mean radius using the mean of the ellipticity profile; 
the central velocity dispersion measurement has been plotted at $r=0.18$ arcsec, corresponding to one half pixel of the spectrograph.)
$B-I$, $I-H$: minor-axis color profiles for the dust-free side of the bulge, derived from combined HST+GB profiles, from Paper\,I.  Abscissa has been projected to the geometric mean radius using the ellipticity profiles.  
$\epsilon$:  {\it triangles} are the ellipticity profile from ellipse fits to the HST/NICMOS F160W images; {\it squares} are the $\epsilon$ profile from ellipse fits to GB $K$-band images, from Peletier \& Balcells (1997); GB $\epsilon$ values are generally kept fixed in the outer, low-S/N region of the images.  
$c4$: fourth-order cosine term of the residuals from ellipse fits ({\it triangles}: HST F160W images; {\it squares}: GB $K$-band images).
}
\label{Fig:MuSigColGeom}
\end{center}
\end{figure*}

\addtocounter{figure}{-1}
\begin{figure*}[htbp]
\begin{center}
\includegraphics[width=\textwidth]{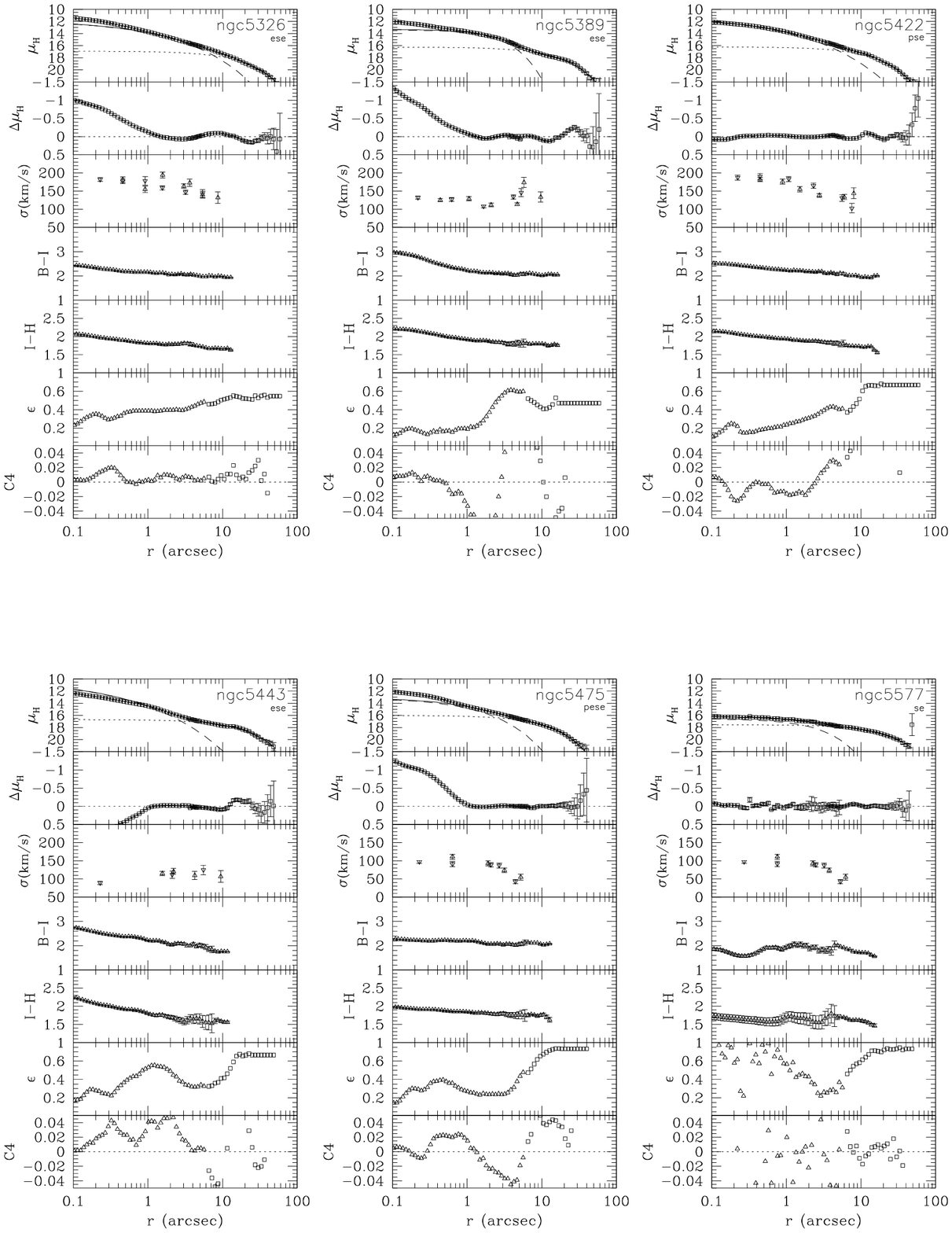}
\caption{See caption in previous page. }
\end{center}
\end{figure*}

\addtocounter{figure}{-1}
\begin{figure*}[htbp]
\begin{center}
\includegraphics[width=\textwidth]{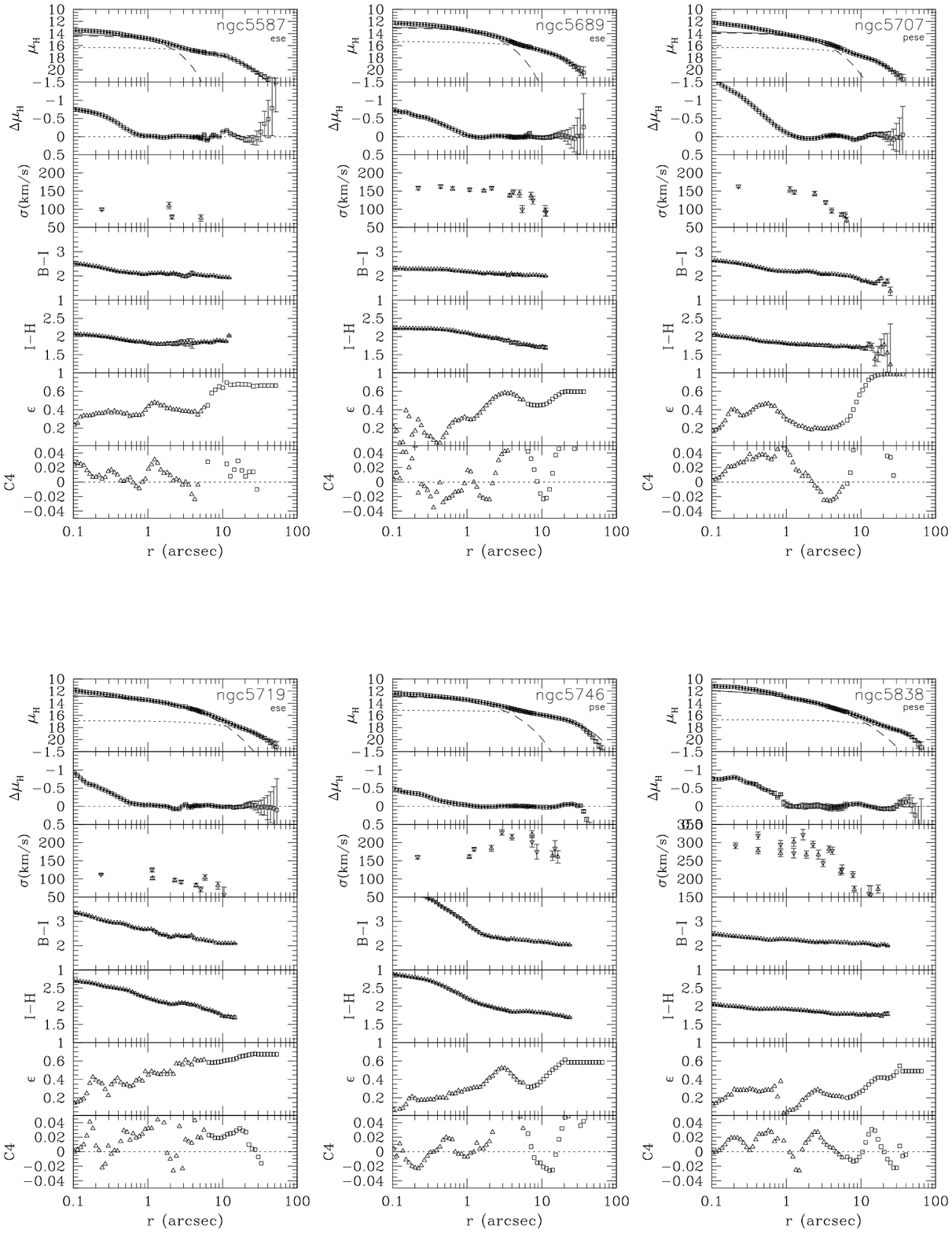}
\caption{continued. }
\end{center}
\end{figure*}

\addtocounter{figure}{-1}
\begin{figure*}[htbp]
\begin{center}
\includegraphics[width=\textwidth]{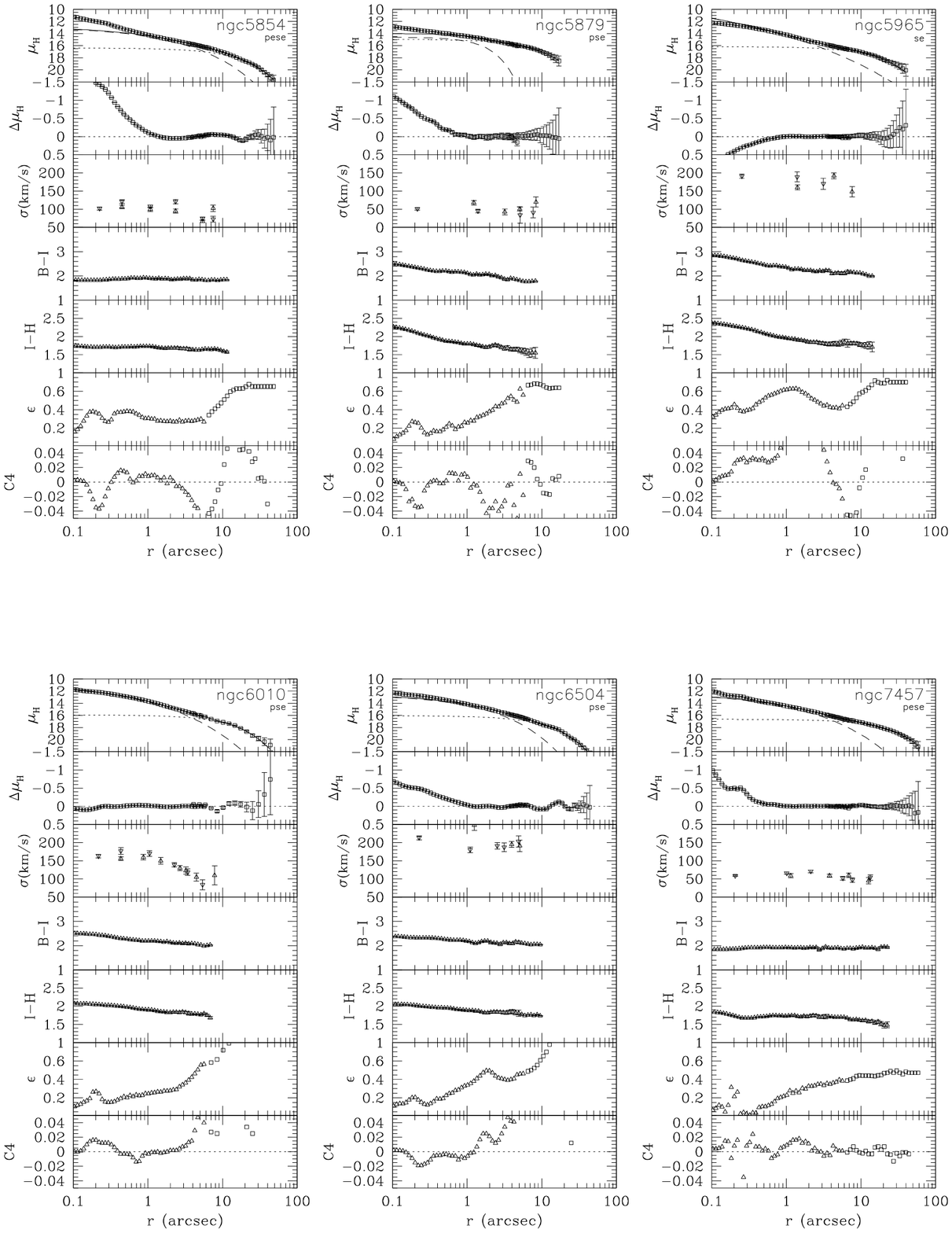}
\caption{continued. }
\end{center}
\end{figure*}

\addtocounter{figure}{-1}
\begin{figure}[htbp]
\begin{center}
\includegraphics[width=0.5\textwidth]{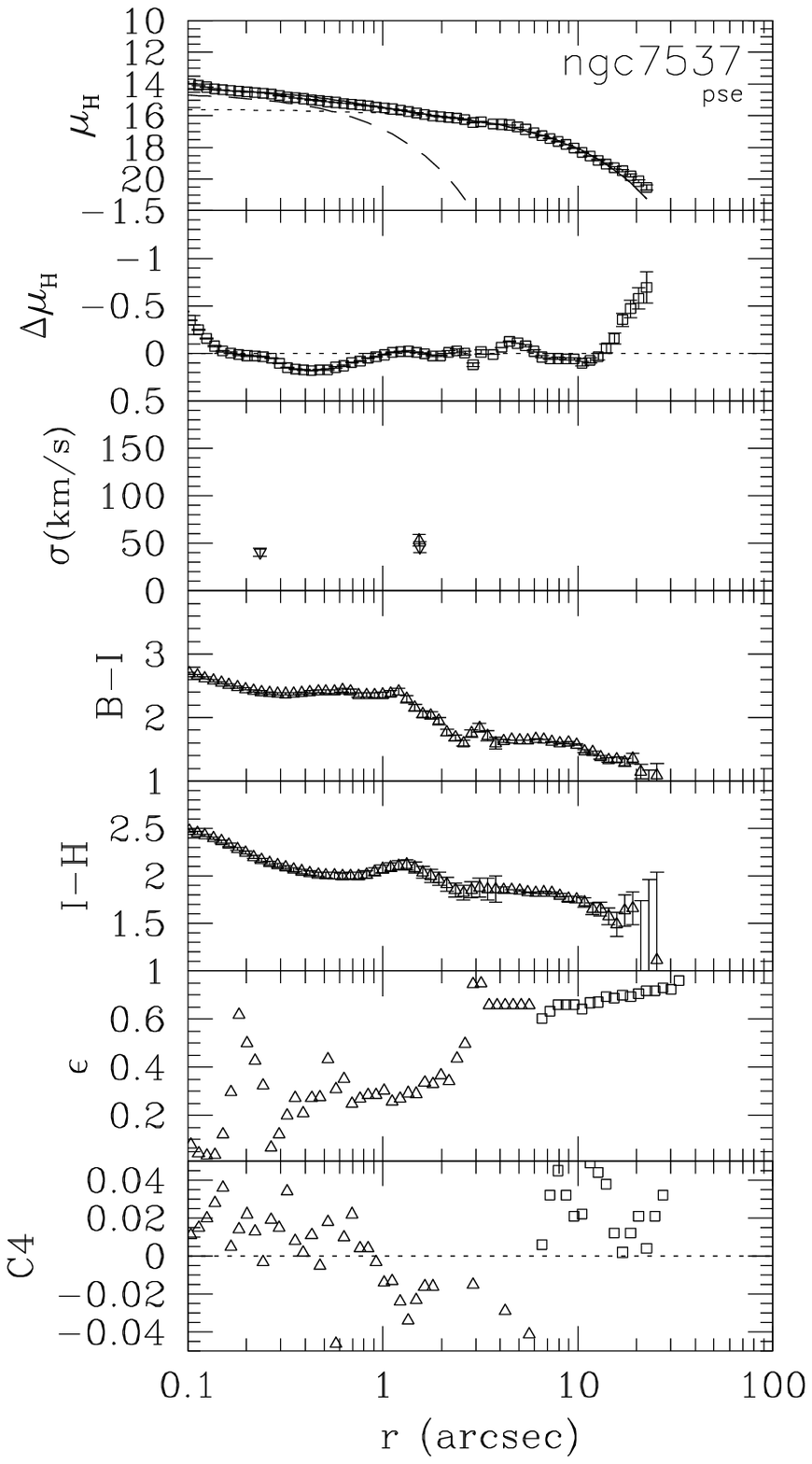}
\caption{continued.  }
\end{center}
\end{figure}

\end{document}

%% file: tabsample.tex
\begin{table*} 
\caption{Galaxy sample. \label{Tab:Sample} }
\begin{center}
\begin{tabular}{cccccccccrcc}
\tableline
NGC & T & $D$ & scale & $K$ & $\pm$ & $M_K$ & $\pm$ & $M_R$ & \multicolumn{1}{c}{$\sigma_0$} & $\pm$ & $\epsilon_\mathrm{Disk}$ \\ 
          &    &  Mpc  & kpc\,arcsec$^{-1}$ & \multicolumn{2}{c}{mag} & \multicolumn{2}{c}{mag} &  mag  & \multicolumn{2}{c}{\kms} &   \\ 
(1) & (2) & (3) & (4) & (5) & (6) & (7) & (8) & (9) & (10) & (11) & (12)  \\ \tableline
5326 & 1 & 34.3 & 0.166 & 8.99 & 0.12 & -23.72 & 0.13 & -21.28 & 164 & 6 & 0.55 \\ 
5389 & 0 & 26.2 & 0.127 & 8.57 & 0.06 & -23.55 & 0.08 & -20.44 & 114 & 6 & 0.75 \\ 
5422 & -2 & 25.3 & 0.123 & 8.88 & 0.05 & -23.16 & 0.08 & -21.08 & 160 & 6 & 0.80 \\ 
5443 & 3 & 27.2 & 0.132 & 9.06 & 0.23 & -23.13 & 0.24 & -20.65 & 76 & 8 & 0.72 \\ 
5475 & 1 & 24.5 & 0.119 & 9.30 & 0.03 & -22.67 & 0.07 & -20.10 & 91 & 6 & 0.71 \\ 
5577 & 4 & 19.6 & 0.095 & 9.53 & 0.08 & -21.96 & 0.11 & -19.63 & ... & ... & 0.72 \\ 
5587 & 0 & 31.0 & 0.150 & 9.61 & 0.09 & -22.88 & 0.10 & -20.19 & 93 & 8 & 0.70 \\ 
5689 & 0 & 30.3 & 0.147 & 8.50 & 0.28 & -23.94 & 0.28 & -21.41 & 143 & 6 & 0.75 \\ 
5707 & 2 & 31.1 & 0.151 & 9.31 & 0.08 & -23.18 & 0.09 & -20.55 & 141 & 6 & 0.75 \\ 
5719 & 2 & 23.1 & 0.112 & 8.40 & 0.09 & -23.44 & 0.11 & -20.72 & 108 & 6 & 0.68 \\ 
5746 & 3 & 22.9 & 0.111 & 6.88 & 0.04 & -24.95 & 0.07 & -21.80 & 139 & 8 & 0.83 \\ 
5838 & -3 & 18.3 & 0.089 & 7.72 & 0.07 & -23.63 & 0.11 & -21.01 & 255 & 6 & 0.63 \\ 
5854 & -1 & 23.3 & 0.113 & 8.63 & 0.17 & -23.24 & 0.18 & -20.55 & 97 & 6 & 0.70 \\ 
5879 & 4 & 13.8 & 0.067 & 8.79 & 0.22 & -21.92 & 0.24 & -19.42 & 58 & 8 & 0.70 \\ 
5965 & 3 & 47.6 & 0.231 & 8.61 & 0.20 & -24.82 & 0.20 & -22.04 & 162 & 8 & 0.83 \\ 
6010 & 0 & 26.0 & 0.126 & 8.82 & 0.34 & -23.31 & 0.34 & -20.69 & 144 & 6 & 0.77 \\ 
6504 & 2 & 61.9 & 0.300 & 9.21 & 0.07 & -24.81 & 0.07 & -23.69 & 185 & 6 & 0.80 \\ 
7457 & -3 & 13.7 & 0.066 & 8.70 & 0.16 & -22.00 & 0.19 & -20.03 & 56 & 6 & 0.48 \\ 
7537 & 4 & 35.3 & 0.171 & 9.68 & 0.20 & -23.11 & 0.20 & -20.49 & 42 & 9 & 0.66 \\ 
\tableline
\end{tabular}
\end{center}
\tablecomments{Column description: (1) Galaxy NGC number.  
(2) Morphological type index from the RC3 (de Vaucouleurs et al.\ 1992). 
(3) Galaxy distance, computed from the Galactic-standard-of-rest recession velocities listed in the RC3, assuming $H_0 = 75$ \kmsMpc.
(4) Spatial scale in kpc\,arcsec$^{-1}$ at the galaxy distance.
(5,6) Galaxy $K$-band apparent magnitude and error, from our photometry (\S~\ref{Sec:Data}). 
(7,8) Galaxy $K$-band absolute magnitude, using the distances in column (2), with Galactic extinction, cosmological correction and K-correction. 
(9) Galaxy $R$-band absolute magnitude, from BP94, for the assumed cosmology.  
(10,11) Aperture-corrected central velocity dispersion and error, from Falc\'on-Barroso et al.\ (2002).
(9) Disk ellipticity derived from $K$-band images, from APB95.}
\end{table*}

%% file: tabfitpars.tex
\begin{table*} 
\caption{Best-fit parameters for the disk, bulge, and nuclear components. \label{Tab:FitParameters} }
\begin{center}
\begin{tabular}{lrrrrrccccrr}
\tableline\tableline
NGC & $\mu_0$ & $h$ & \mueff\ & \reff\ & $n$ & $B/D$ & $H_\mathrm{PS}$ & $H_\mathrm{E2}$ & $\mu_{0,2}$ & $h_2$ & Fit type \\ 
(1) & (2) & (3) & (4) & (5) & (6) & (7) & (8) & (9) & (10) & (11) & (12) \\ \tableline
5326 & 17.10 &   11.47 & 15.93 & 3.88 & 2.60 &   0.99 & ... &  13.64 & 11.55 & 0.15 & ese \\ 
5389 & 16.18 &    9.70 & 14.90 & 2.29 & 1.35 &   0.39 & ... &  15.19 & 11.04 & 0.06 & ese \\ 
5422 & 16.11 &    8.61 & 15.34 & 2.56 & 2.04 &   0.47 & 17.79 &    ... & ... & ... & pse \\ 
5443 & 16.67 &   11.56 & 15.28 & 1.41 & 1.69 &   0.13 & ... &  16.34 & 10.57 & 0.03 & ese \\ 
5475 & 16.15 &    7.27 & 16.32 & 2.52 & 2.12 &   0.27 & 17.51 &  13.50 & 12.24 & 0.22 & pese \\ 
5577 & 17.50 &   13.30 & 18.45 & 2.75 & 1.09 &   0.04 & ... &    ... & ... & ... & se \\ 
5587 & 16.28 &    7.63 & 15.69 & 1.43 & 0.71 &   0.10 & ... &  14.49 & 13.59 & 0.26 & ese \\ 
5689 & 15.33 &    7.38 & 14.54 & 2.03 & 0.86 &   0.28 & ... &  13.39 & 12.48 & 0.26 & ese \\ 
5707 & 16.25 &    7.41 & 15.53 & 2.72 & 0.78 &   0.45 & 16.87 &  12.95 & 12.10 & 0.27 & pese \\ 
5719 & 17.30 &   14.83 & 15.72 & 5.48 & 1.96 &   1.50 & ... &  15.34 & 10.06 & 0.04 & ese \\ 
5746 & 15.24 &   14.63 & 15.15 & 2.98 & 1.55 &   0.10 & 17.65 &    ... & ... & ... & pse \\ 
5838 & 16.54 &   17.24 & 15.14 & 4.82 & 1.44 &   0.63 & 15.76 &  11.42 & 11.21 & 0.36 & pese \\ 
5854 & 16.15 &    9.30 & 16.44 & 3.94 & 1.88 &   0.35 & 15.45 &  12.96 & 11.02 & 0.16 & pese \\ 
5879 & 15.08 &    5.48 & 16.54 & 2.18 & 2.23 &   0.11 & 17.97 &    ... & ... & ... & pse \\ 
5965 & 16.21 &   10.99 & 16.08 & 2.68 & 2.75 &   0.20 & ... &    ... & ... & ... & se \\ 
6010 & 15.81 &    8.08 & 15.02 & 2.04 & 2.12 &   0.36 & 16.89 &    ... & ... & ... & pse \\ 
6504 & 16.22 &    7.55 & 16.32 & 3.54 & 2.65 &   0.59 & 17.85 &    ... & ... & ... & pse \\ 
7457 & 16.61 &   14.50 & 16.45 & 3.41 & 1.99 &   0.17 & 15.53 &  14.55 & 12.53 & 0.16 & pese \\ 
7537 & 15.64 &    4.44 & 17.14 & 1.12 & 1.76 &   0.04 & 17.99 &    ... & ... & ... & pse \\ 
\tableline
\end{tabular}
\end{center}
\tablecomments{Table lists output parameters from the
surface brightnesss profile fitting code, prior to applying the corrections
described in \S~\ref{Sec:ParameterUncertainties}.   
All surface brightnesses are given in $H$-band mag\,arcsec$^{-2}$, and are not corrected to face-on view. 
All scale-lengths, in arcsec, refer to the geometric mean axis $\sqrt(a \times b)$ of each measured ellipse.
(1) Galaxy NGC number.    
(2) Extrapolated disk central surface brightness.  (3) Disk scale length.  
(4) Bulge effective surface brightness.  (5) Bulge effective radius.  (6) Bulge S\'ersic index.
(7) Luminosity ratio between bulge and main disk, from best-fit parameters ($H$-band). 
(8) $H$-band magnitude of central unresolved source, from  best-fit parameters ($H$-band). 
(9) $H$-band magnitude of nuclear exponential component, from best-fit parameters ($H$-band).  
(10) Extrapolated central surface brightness of nuclear exponential.  
(11) Scale length of nuclear exponential.   
(12) Fit type code.  \textit{se}: S\'ersic bulge and exponential outer disk.  
     \textit{pse}: \textit{se} plus a nuclear point source. 
     \textit{ese}: \textit{se} plus an inner exponential component. 
     \textit{pese}: \textit{se} plus a nuclear point source and a nuclear exponential component. }
\end{table*}

%% file: tabmainphyspars.tex
\begin{table*} 
\caption{Physical parameters for the disk and bulge components. \label{Tab:MainPhysParameters} }
\begin{center}
\begin{tabular}{lcccccccrcccccrc}
\tableline\tableline
NGC  & $M_{K,\mathrm{Bulge}}$ & $\pm$ & $M_{K,\mathrm{Disk}}$ &  $\pm$ & $B/D$ & $\mu_0$ & $\pm$ & \multicolumn{1}{c}{$\log(h)$} & $\pm$ & \mueff\ & $\pm$ & $\log(\reff)$ & $\pm$ & \multicolumn{1}{c}{$\log(n)$} & $\pm$ \\ 
(1) & (2) & (3) & (4) & (5) & (6) & (7) & (8) & (9) & \multicolumn{1}{c}{(10)} & (11) & (12) & (13) & (14)  & \multicolumn{1}{c}{(15)} & (16) \\ \tableline
5326 &  -22.96 & 0.17 & -22.98 & 0.14 & 0.98 & 17.94 &   0.04 & 0.455 &  0.004 & 15.69 & 0.35 & -0.23 & 0.07 & 0.41 & 0.07 \\ 
5389 &  -22.17 & 0.14 & -23.19 & 0.09 & 0.39 & 17.66 &   0.04 & 0.392 &  0.004 & 14.66 & 0.35 & -0.58 & 0.07 & 0.13 & 0.07 \\ 
5422 &  -21.86 & 0.10 & -22.77 & 0.13 & 0.43 & 17.84 &   0.12 & 0.374 &  0.014 & 15.26 & 0.22 & -0.53 & 0.05 & 0.31 & 0.04 \\ 
5443 &  -20.76 & 0.26 & -23.00 & 0.24 & 0.13 & 18.03 &   0.04 & 0.460 &  0.004 & 15.06 & 0.35 & -0.77 & 0.07 & 0.23 & 0.07 \\ 
5475 &  -20.98 & 0.14 & -22.41 & 0.08 & 0.27 & 17.47 &   0.04 & 0.206 &  0.004 & 16.10 & 0.35 & -0.57 & 0.07 & 0.33 & 0.07 \\ 
5577 &  -18.20 & 0.13 & -21.93 & 0.15 & 0.03 & 18.85 &   0.12 & 0.378 &  0.014 & 18.36 & 0.22 & -0.61 & 0.05 & 0.04 & 0.04 \\ 
5587 &  -20.26 & 0.16 & -22.78 & 0.11 & 0.10 & 17.56 &   0.04 & 0.321 &  0.004 & 15.46 & 0.35 & -0.71 & 0.07 & -0.15 & 0.07 \\ 
5689 &  -22.27 & 0.31 & -23.68 & 0.29 & 0.27 & 16.80 &   0.04 & 0.337 &  0.004 & 14.31 & 0.35 & -0.57 & 0.07 & -0.07 & 0.07 \\ 
5707 &  -21.90 & 0.15 & -22.78 & 0.10 & 0.44 & 17.73 &   0.04 & 0.350 &  0.004 & 15.30 & 0.35 & -0.43 & 0.07 & -0.11 & 0.07 \\ 
5719 &  -22.88 & 0.16 & -22.45 & 0.12 & 1.49 & 18.51 &   0.04 & 0.467 &  0.004 & 15.48 & 0.35 & -0.26 & 0.07 & 0.29 & 0.07 \\ 
5746 &  -22.31 & 0.10 & -24.85 & 0.13 & 0.10 & 17.13 &   0.12 & 0.596 &  0.014 & 15.06 & 0.22 & -0.51 & 0.05 & 0.19 & 0.04 \\ 
5838 &  -22.59 & 0.16 & -23.10 & 0.12 & 0.62 & 17.59 &   0.04 & 0.401 &  0.004 & 14.91 & 0.35 & -0.41 & 0.07 & 0.16 & 0.07 \\ 
5854 &  -21.75 & 0.22 & -22.92 & 0.19 & 0.34 & 17.42 &   0.04 & 0.282 &  0.004 & 16.20 & 0.35 & -0.40 & 0.07 & 0.27 & 0.07 \\ 
5879 &  -19.35 & 0.25 & -21.81 & 0.26 & 0.10 & 16.38 &   0.12 & -0.174 &  0.014 & 16.47 & 0.22 & -0.87 & 0.05 & 0.35 & 0.04 \\ 
5965 &  -22.81 & 0.21 & -24.63 & 0.23 & 0.19 & 18.10 &   0.12 & 0.789 &  0.014 & 15.98 & 0.22 & -0.24 & 0.05 & 0.44 & 0.04 \\ 
6010 &  -21.79 & 0.35 & -23.00 & 0.36 & 0.33 & 17.35 &   0.12 & 0.328 &  0.014 & 14.90 & 0.22 & -0.62 & 0.05 & 0.33 & 0.04 \\ 
6504 &  -23.68 & 0.10 & -24.34 & 0.13 & 0.55 & 17.91 &   0.12 & 0.704 &  0.014 & 16.20 & 0.22 & -0.00 & 0.05 & 0.42 & 0.04 \\ 
7457 &  -19.87 & 0.23 & -21.84 & 0.20 & 0.16 & 17.29 &   0.04 & 0.125 &  0.004 & 16.22 & 0.35 & -0.69 & 0.07 & 0.30 & 0.07 \\ 
7537 &  -19.47 & 0.22 & -23.07 & 0.23 & 0.04 & 16.76 &   0.12 & 0.114 &  0.014 & 17.03 & 0.22 & -0.75 & 0.05 & 0.24 & 0.04 \\ 
\tableline
\end{tabular}
\end{center}
\tablecomments{Table lists parameters for disk and bulge 
corrected from measurement offsets as described in \S~\ref{Sec:ParameterUncertainties}.  
Absolute magnitudes are given in the $K$-band.    
Surface brightnesses are given in $H$-band mag\,arcsec$^{-2}$.
Both magnitudes and surface brightnesses are corrected for Galactic extinction, cosmological dimming, and K-correction.   
Scale-lengths are in kpc;   
for bulges, they refer to the geometric mean axis $(ab)^{1/2}$ of the measured ellipse, while, 
for disks, scale-lengths are scaled to the major axis, assuming an inclination given by the disk ellipticity.  
(1) Galaxy NGC number.    
(2-5) Bulge and disk $K$-band absolute magnitudes and errors, from the galaxy $K$-band absolute magnitude and the bulge-disk ratio from col.~(6).
(6) Luminosity ratio between bulge and main disk, from best-fit parameters ($H$-band). 
(7,8) Disk face-on extrapolated central surface brightness, and error.  
(9,10) Disk scale length, and error.  
(11,12) Bulge effective surface brightness, and error.  
(13,14) Bulge effective radius, and error.  
(15,16) Bulge S\'ersic index, and error.}
\end{table*}

%% file: tabnucphyspars.tex
\begin{table*} 
\caption{Physical parameters for the nuclear components. \label{Tab:NucPhysParameters} }
\begin{center}
\begin{tabular}{lccccccccr}
\tableline\tableline
NGC & $M_{K,\mathrm{PS}}$ & $\pm$ & $M_{K,\mathrm{E2}}$ & $\pm$ & $\mu_{0,2}$ & $\pm$ & \multicolumn{1}{c}{$\log(h_2)$} & $\pm$ & Fit type \\ 
(1) & (2) & (3) & (4) & (5) & (6) & (7) & (8) & (9) & (10) \\ \tableline
5326 & ... &  ... & -19.21 & 0.47 & 11.58 &   0.94 & -1.60 &   0.18 & ese \\ 
5389 & ... &  ... & -17.07 & 0.47 & 11.71 &   0.94 & -2.00 &   0.18 & ese \\ 
5422 & -14.79 &  0.35 & ... & ... & ... &    ... & ... &    ... & pse \\ 
5443 & ... &  ... & -15.99 & 0.47 & 11.12 &   0.94 & -2.34 &   0.18 & ese \\ 
5475 & -15.00 &  0.35 & -18.70 & 0.26 & 13.11 &   0.45 & -1.40 &   0.04 & pese \\ 
5577 & ... &  ... & ... & ... & ... &    ... & ... &    ... & se \\ 
5587 & ... &  ... & -18.23 & 0.26 & 14.42 &   0.45 & -1.23 &   0.04 & ese \\ 
5689 & ... &  ... & -19.28 & 0.26 & 13.50 &   0.45 & -1.20 &   0.04 & ese \\ 
5707 & -16.16 &  0.35 & -19.77 & 0.26 & 13.13 &   0.45 & -1.18 &   0.04 & pese \\ 
5719 & ... &  ... & -16.64 & 0.47 & 10.45 &   0.94 & -2.34 &   0.18 & ese \\ 
5746 & -14.72 &  0.35 & ... & ... & ... &    ... & ... &    ... & pse \\ 
5838 & -16.13 &  0.35 & -20.16 & 0.26 & 11.81 &   0.45 & -1.36 &   0.04 & pese \\ 
5854 & -16.96 &  0.35 & -19.14 & 0.26 & 11.84 &   0.45 & -1.56 &   0.04 & pese \\ 
5879 & -13.28 &  0.35 & ... & ... & ... &    ... & ... &    ... & pse \\ 
5965 & ... &  ... & ... & ... & ... &    ... & ... &    ... & se \\ 
6010 & -15.78 &  0.35 & ... & ... & ... &    ... & ... &    ... & pse \\ 
6504 & -16.71 &  0.35 & ... & ... & ... &    ... & ... &    ... & pse \\ 
7457 & -15.71 &  0.35 & -16.29 & 0.47 & 12.40 &   0.94 & -2.02 &   0.18 & pese \\ 
7537 & -15.34 &  0.35 & ... & ... & ... &    ... & ... &    ... & pse \\ 
\tableline
\end{tabular}
\end{center}
\tablecomments{Table lists parameters for nuclear components
corrected from measurement offsets as described in \S~\ref{Sec:ParameterUncertainties}.  
Nuclear source absolute magnitudes are given in the $K$-band,  
using $H-K=0.23$.  
Surface brightness $\mu_{0,2}$ is given in $H$-band mag\,arcsec$^{-2}$.
Both magnitudes and surface brightnesses are corrected for Galactic extinction, cosmological dimming, and K-correction.   
Scale-lengths are in kpc, and have been scaled to the major axis, 
assuming an inclination given by the disk ellipticity.  
(1) Galaxy NGC number.    
(2,3) $K$-band absolute magnitude of the central unresolved source, and error. 
(4,5) $K$-band absolute magnitude of the nuclear exponential component, and error.
(6,7) Face-on extrapolated central surface brightness, and error, of the nuclear exponential component.  
(8,9) Scale length, and error, of the nuclear exponential component.  
(10) Fit type code.  \textit{se}: S\'ersic bulge and exponential outer disk.  
     \textit{pse}: \textit{se} plus a nuclear point source. 
     \textit{ese}: \textit{se} plus an inner exponential component. 
     \textit{pese}: \textit{se} plus a nuclear point source and a nuclear exponential component. }
\end{table*}

%% file: tabgammas.tex
\begin{table} 
\caption{Inner negative logarithmic slopes. \label{Tab:Gammas} }
\begin{center}
\begin{tabular}{cccccc}
\tableline
\tableline
NGC & $\gamma$(pwl) & $\gamma$(pwl) & $\gammaNukPrime$ &  $\gammaSerPrime$ & $\gammaSerPrime$  \\ 
    & 0.1-1$\arcsec$ & 20-200 pc      &   0.3$\arcsec$   &      0.3$\arcsec$ &    0.15\reff\       \\    
(1) & (2) & (3) & (4) & (5) & (6) \\ \tableline
5326 & 0.91 & 0.85 & 0.98 & 0.90 & 0.70 \\ 
5389 & 0.69 & 0.70 & 0.72 & 0.43 & 0.39 \\ 
5422 & 0.62 & 0.65 & 0.63 & 0.72 & 0.64 \\ 
5443 & 0.83 & 0.83 & 0.84 & 0.59 & 0.72 \\ 
5475 & 0.96 & 1.03 & 1.34 & 0.75 & 0.68 \\ 
5577 & 0.17 & 0.22 & 0.16 & 0.30 & 0.22 \\ 
5587 & 0.56 & 0.52 & 0.81 & 0.11 & 0.17 \\ 
5689 & 0.46 & 0.43 & 0.49 & 0.18 & 0.17 \\ 
5707 & 0.79 & 0.77 & 0.86 & 0.14 & 0.10 \\ 
5719 & 0.66 & 0.65 & 0.67 & 0.70 & 0.42 \\ 
5746 & 0.43 & 0.46 & 0.46 & 0.53 & 0.41 \\ 
5838 & 0.67 & 0.84 & 0.78 & 0.47 & 0.26 \\ 
5854 & 1.24 & 1.26 & 1.21 & 0.67 & 0.46 \\ 
5879 & 0.63 & 0.67 & 0.65 & 0.79 & 0.76 \\ 
5965 & 0.78 & 0.61 & 0.84 & 0.94 & 0.85 \\ 
6010 & 0.74 & 0.77 & 0.76 & 0.75 & 0.75 \\ 
6504 & 0.73 & 0.54 & 0.78 & 0.92 & 0.74 \\ 
7457 & 0.96 & 0.97 & 0.92 & 1.12 & 0.82 \\ 
7537 & 0.59 & 0.63 & 0.59 & 0.62 & 0.86 \\ 
\tableline
\end{tabular}
\end{center}
\tablecomments{Column description: (1) galaxy NGC number.  
(2) $\gamma$ from power-law fit to 0.1\arcsec--1\arcsec (Fig.~\ref{Fig:InnerProfiles}).
(3) $\gamma$ from power-law fit to 20\,pc--200\,pc.
(4) $\gammaNukPrime$, slope of Nuker model at $ R = 0.3\arcsec$
(single Nuker fit to $0.1\arcsec < R \la 4\arcsec$).
(5) $\gammaSerPrime$, slope of S\'ersic model at $ R = 0.3\arcsec$  
(simultaneous bulge, disk and nuclear component fits).
(6) $\gammaSerPrime$, slope of S\'ersic model at $ R /\reff = 0.15$.}
\end{table}